# Tolerance analysis by polytopes: taking into account degrees of freedom with cap half-spaces

Lazhar HOMRI
l.homri@i2m.u-bordeaux1.fr

Denis TEISSANDIER
d.teissandier@i2m.u-bordeaux1.fr

Alex BALLU
a.ballu@i2m.u-bordeaux1.fr

Univ. Bordeaux, I2M, UMR 5295, F-33400 Talence, France.

***Abstract:***

*To determine the relative position of any two surfaces in a system, one approach is to use operations (Minkowski sum and intersection) on sets of constraints. These constraints are made compliant with half-spaces of $\mathbb{R}^n$ where each set of half-spaces defines an operand polyhedron. These operands are generally unbounded due to the inclusion of degrees of invariance for surfaces and degrees of freedom for joints defining theoretically unlimited displacements. To solve operations on operands, Minkowski sums in particular, "cap" half-spaces are added to each polyhedron to make it compliant with a polytope which is by definition a bounded polyhedron. The difficulty of this method lies in controlling the influence of these additional half-spaces on the topology of polytopes calculated by sum or intersection. This is necessary to validate the geometric tolerances that ensure the compliance of a mechanical system in terms of functional requirements.*

**Keywords:** Tolerance analysis, degree of freedom, polyhedron, polytope, cap half-spaces, Minkowski sum, intersection.

# 1 Introduction

When analyzing the geometric tolerances of a mechanical system, the traditional approach is to handle sets of linear constraints [1], [2], [3], [4], [5], [6] and [7]. These sets of constraints define the boundaries of relative displacements between two surfaces of the same part (geometric constraints) and boundaries of relative displacements between two surfaces of two separate parts, but which are potentially in contact (contact constraints).

Sets of geometric and contact constraints are generally operand sets which can be made compliant with finite sets of half-spaces of $\mathbb{R}^n$ [4]. The boundaries of relative displacements between two surfaces result from the intersection of the half-spaces of $\mathbb{R}^n$ of an operand set, defining an operand polyhedron. A polyhedron is not usually bounded, due to the degree of invariance of a surface or to the degree of freedom of joints defining theoretically unlimited displacements [8], [9].

The relative position between any two surfaces of a mechanism is determined by operations on these operand polyhedrons (Minkowski sum and intersection), [9], [10] and [11]. To solve these operations, and in particular the Minkowski sums in $\mathbb{R}^n$, one method is to delimit the intersection of the half-spaces of an operand set and thus transform a polyhedron of $\mathbb{R}^n$ into a polytope of $\mathbb{R}^n$, as a polytope of $\mathbb{R}^n$ is a bounded polyhedron of $\mathbb{R}^n$ [12]. This method is justified by the algorithmic complexity of summing polyhedra of $\mathbb{R}^n$, requiring the development of algorithms to compute tolerance analysis of the Minkowski sums of polytopes [13], [14] and [15].

This article introduces the concept of "cap" half-spaces to delimit operand polyhedra. They are added to the operand set and in this way determining the relative position of two surfaces of a mechanical systems is based solely on operations on operand polytopes generating a calculated polytope [16].

By checking that a calculated polytope is included within a functional polytope the conformity of a mechanical system can be simulated with respect to a functional requirement [4], see Fig. 1.

The addition of cap half-spaces to the operand sets will affect the topology of a calculated polytope. Hence it has to be possible to differentiate among all the facets of a calculated polytope between those that are generated by the cap half-spaces and the others generated by half-spaces that derive from geometric and contact constraints. This is essential in order to validate the geometric tolerances that ensure that a mechanical system is compliant in relation to a functional requirement.

This article describes how to identify the facets generated by the cap half-spaces of a polytope resulting from a Minkowski sum or an intersection between two operand polytopes.

In the following, we limit ourselves to 6-dimension polyhedra and polytopes: the half-spaces arising from the geometric and contact constraints are linear inequalities in six variables: three rotation variables and three translation variables [16].

In the first part, some properties of polyhedra and polytopes are considered; the second part looks at determining the cap half-spaces which set boundaries to the half-space intersections resulting from geometric and contact constraints.

The third part deals with the two methods of identifying the dependent facets of cap half-spaces in a summation and an intersection respectively. An example of an application of tolerance analysis illustrating these two methods is described at the end of the article.

For the application used as an example, we put forward the following physical hypotheses:

- no form defect in the real surfaces,
- no local strain in surfaces in contact,
- no deformable parts.

# 2 Some definitions and properties of polyhedra and polytopes

We first set out some definitions and properties of polyhedra and polytopes to ensure a proper understanding of the rest of the article. These are taken from [12].

## 2.1 Hyperplane, half-space

A hyperplane is an affine subspace of dimension 0, 1, 2 or $(n-1)$ in $\mathbb{R}^n$ called point, line, plane and hyperplane respectively. A hyperplane of dimension $(n-1)$ is denoted a $(n-1)$-hyperplane.

Let us consider $H$ a hyperplane in some $\mathbb{R}^n$ : $a_1x_1 + ... + a_ix_i + ... + a_nx_n = b$ .

$$H = \left\{ \mathbf{x} \in \mathbb{R}^n : \ \mathbf{a}^T\mathbf{x} = b \right\} \text{ with } \mathbf{a}^T = (a_1, \ldots a_n) \in \mathbb{R}^n, \mathbf{x} = (x_1, \ldots x_n)^T \in \mathbb{R}^n \text{ and } b \in \mathbb{R} \quad (1)$$

$\overline{H}^+$ is the closed half-space: $a_1x_1 + ... + a_ix_i + ... + a_nx_n \geq b$

$\overline{H}^-$ is the closed half-space: $a_1x_1 + ... + a_ix_i + ... + a_nx_n \leq b$

In the following any manipulated constraint will be written in terms of $\overline{\mathrm{H}}^-$ denoted simply:

$$\overline{H}^- = \left\{ \mathbf{x} \in \mathbb{R}^n : \ \mathbf{a}^T\mathbf{x} \leq b \right\} \quad (2)$$

## 2.2 Polyhedron, polytope

A polyhedron $\mathcal{P}$ is the intersection of a finite number of closed half-spaces of $\mathbb{R}^n$ . We define:

$$\begin{aligned} \mathcal{P} &= \bigcap_{i=1}^{m} \overline{H_i}^- = \left\{ \mathbf{x} \in \mathbb{R}^n : \ \mathbf{a_i}^T\mathbf{x} \leq b_i \ , \ i = 1,...,\mathrm{m} \right\} \\ &= \left\{ \mathbf{x} \in \mathbb{R}^n : \ \mathbf{A}\mathbf{x} \leq \mathbf{b} \ \right\}, \mathbf{A} \in \mathbb{R}^{m \times n} \end{aligned} \quad (3)$$

Where $\mathbf{a_i}^T$ is the i$^{\text{th}}$ line of $\mathbf{A}$ and $b_i$ the i$^{\text{th}}$ component of $\mathbf{b}$.

This definition characterizes the $\mathcal{H}$ – description of $\mathcal{P}$ , see Fig. 2(a), [9] and [18].

A hyperplane $H$ is said to be a support hyperplane for polyhedron $\mathcal{P}$ if and only if:

$$\mathcal{P} \cap H \neq \varnothing \text{ and } \mathcal{P} \subset \overline{H}^- \quad (4)$$

A face $F$ of $\mathcal{P}$ is the intersection of the polyhedron $\mathcal{P}$ with one of its support hyperplanes. The faces of a $d$ -polyhedron $\mathcal{P}$ are convex sub-sets of dimension $k$, $0 \leq k \leq d-1$.

A face of dimension $d$ is called $d$ -face. A 0-face is a vertex, a 1-face is an edge and a ($d-1$ )-face is a facet of $\mathcal{P}$.

If the polyhedron $\mathcal{P}$ is bounded, it is called a polytope. A polytope of dimension $d$ is denoted $d$ – polytope in $\mathbb{R}^n$ ( $n \geq d$ ), see Fig. 2(b).

By duality, a polytope $\mathcal{P}$ can be defined by the convex hull of a finite number of points of $\mathbb{R}^n$.

Let us consider $V$ a finite set of points of $\mathbb{R}^n$:

$$\mathcal{P} = \text{conv}(V) \tag{5}$$

Definition (5) is the $\mathcal{V}$ – description of $\mathcal{P}$ [9] and [18].

Hence a polytope can be defined by its vertices.

In this article we consider only convex polyhedra and polytopes, which we call simply polyhedra and polytopes.

## 2.3 Primal cone, dual cone and normal fan

Associated to each vertex $v$ of polytope $\mathcal{P}$, is a primal cone and a dual cone such that:

- the primal cone consists of the set of edges and facets of the polytope derived from this vertex (see Fig. 2(c)),
- the dual (or normal) cone is constructed by taking into account the following properties (see Fig. 2(c)):
  - Each edge of the primal cone is normal to a facet of the dual cone,
  - Each facet of the primal cone is orthogonal to an edge of the dual cone.

A fan of $\mathbb{R}^n$ is a family $\mathcal{F} = \{C_1, C_2, ..., C_k\}$ of cones such that:

- each non-empty face of a cone of $\mathcal{F}$ is also a cone of $\mathcal{F}$,
- the intersection of two cones of $\mathcal{F}$ is a face common to the two cones.

Fan $\mathcal{F}$ is complete if $\bigcup_{i=1}^{k} C_i = \mathbb{R}^n$.

If $C_1, C_2, ..., C_k$ are the dual cones of a polytope $\mathcal{P}$, then $\mathcal{F}$ is the normal fan of polytope $\mathcal{P}$ denoted $N(\mathcal{P})$.

Fig. 2(d) and Fig. 2(e) illustrate the normal fans for a polyhedron and a polytope respectively.

The common refinement of two fans is defined by the intersection of pairs of cones from the two families [17]:

$$\mathcal{F}_1 \wedge \mathcal{F}_2 = \{C_1 \cap C2 \;:\; C_1 \in \mathcal{F}_1, C_2 \in \mathcal{F}_2\} \tag{6}$$

# 3 Operand polytopes in geometric tolerancing

## 3.1 Geometric constraints

Whatever manufacturing processes are employed, and whatever materials are used, a manufactured part is never geometrically perfect. The geometric features of a manufactured part will never be exactly the same as the nominal characteristics of which they are a specific physical realization. Generally, variations in production processes generate geometric deviations from a geometric model produced by a Computer Assisted Design tool (CAD). These deviations, also called defects, may compromise the expected performance of a mechanical system. Tolerancing sets an acceptable range of deviation using geometric specifications while still guaranteeing that the mechanical system will function as required. These limits are mainly given in the form of tolerance zones, or zones within which the surfaces being produced must be included.

In mathematical terms, the displacement limits for a surface in a tolerance zone are defined by a set of geometric constraints. A tolerance zone is defined by its dimension $t$ and two deviations $d^{\inf}$ and $d^{\sup}$ ($d^{\sup}-d^{\inf}=t$ with $d^{\inf}\le d^{\sup}$) which position it in relation to the nominal surface $S_0$. Each real surface is modeled by a substitution surface (i.e. one which is perfect) with $S_1$ representing a specific physical representation of the nominal surface $S_0$ (see Fig. 3). If $|d^{\inf}|=|d^{\sup}|=\frac{t}{2}$, the nominal surface is combined with the median surface of the tolerance zone.

The set of geometric constraints is defined in the following equation [4]:

$$S_1 \subseteq ZT \Leftrightarrow \forall N_i \in S_0 : d^{\inf} \le \mathbf{t}_{N_i}.\mathbf{n}_i \le d^{\sup} \tag{7}$$

Where $\mathbf{t}_{N_i}$ is the translation displacement of $S_1$ in relation to $S_0$ at point $N_i$, $\mathbf{n}_i$ is the unit outward pointing vector normal to the surface $S_0$ in $N_i$.

By discretizing the surface $S_0$ into $n$ points $N_i$ gives $n$ equations (7) with $1\le i\le n$.

Using small displacement torsors, equation (8) can be written for every point $M$ of the Euclidean space [19]:

$$\mathbf{t}_{Ni} = \mathbf{t}_M + \mathbf{N}_i\mathbf{M}\times\mathbf{r} \tag{8}$$

Where:

$\mathbf{r}$ : rotation vector of $S_1$ in relation to $S_0$,

$\mathbf{t}_M$ : translation vector of $S_1$ in relation to $S_0$ at point $M$ .

Developing equation (8) gives (9):

$$\mathbf{t}_{Ni}.\mathbf{n}_i = \left(\mathbf{t}_M + \mathbf{N}_i\mathbf{M}\times\mathbf{r}\right).\mathbf{n}_i = \mathbf{n}_i.\mathbf{t}_M + \left(\mathbf{MN}_i\times\mathbf{n}_i\right).\mathbf{r} = \begin{pmatrix} \mathbf{MN}_i\times\mathbf{n}_i \\ \mathbf{n}_i \end{pmatrix}^T \begin{pmatrix} \mathbf{r} \\ \mathbf{t}_M \end{pmatrix} \tag{9}$$

We define the components of the following vectors in an $(\mathbf{x},\mathbf{y},\mathbf{z})$ base:

$\mathbf{t}_M\begin{pmatrix} t_{Mx} \\ t_{My} \\ t_{Mz} \end{pmatrix}$, $\mathbf{N}_i\begin{pmatrix} x_i \\ y_i \\ z_i \end{pmatrix}$, $\mathbf{M}\begin{pmatrix} x_M \\ y_M \\ z_M \end{pmatrix}$, $\mathbf{n}_i\begin{pmatrix} n_{ix} \\ n_{iy} \\ n_{iz} \end{pmatrix}$ and $\mathbf{r}\begin{pmatrix} r_x \\ r_y \\ r_z \end{pmatrix}$

Thus equation (9) can be written in the $(\mathbf{x},\mathbf{y},\mathbf{z})$ base in accordance with (10):

$$\mathbf{t}_{N_i}.\mathbf{n}_i = \left(\mathbf{MN}_i\times\mathbf{n}_i\right).\mathbf{r} + \mathbf{n}_i.\mathbf{t}_M = \begin{pmatrix} n_{iz}\left(y_i-y_M\right)-n_{iy}\left(z_i-z_M\right) \\ n_{ix}\left(z_i-z_M\right)-n_{iz}\left(x_i-x_M\right) \\ n_{iy}\left(x_i-x_M\right)-n_{ix}\left(y_i-y_M\right) \\ n_{ix} \\ n_{iy} \\ n_{iz} \end{pmatrix}^T \begin{pmatrix} r_x \\ r_y \\ r_z \\ t_{Mx} \\ t_{My} \\ t_{Mz} \end{pmatrix} = a_i^T\mathbf{x} \tag{10}$$

We propose:

$$\mathbf{A}_1 = \begin{pmatrix} \mathbf{MN}_1 \times \mathbf{n}_1 & \cdots & \mathbf{MN}_i \times \mathbf{n}_i & \cdots & \mathbf{MN}_n \times \mathbf{n}_n \\ \mathbf{n}_1 & \cdots & \mathbf{n}_i & \cdots & \mathbf{n}_n \end{pmatrix}^T = \begin{pmatrix} a_1 & \cdots & a_{\mathbf{i}} & \cdots & a_n \end{pmatrix}^T \in \mathbb{R}^{n\times 6},$$

$$\mathbf{A}_2 = -\mathbf{A}_1,$$

$$\mathbf{x} = \begin{pmatrix} \mathbf{r} \\ \mathbf{t}_M \end{pmatrix} \in \mathbb{R}^6, \tag{11}$$

$$\mathbf{b}_1 = \left(d^{\sup}\right)_n \in \mathbb{R}^n \text{ and } \mathbf{b}_2 = \left(-d^{\inf}\right)_n \in \mathbb{R}^n.$$

Finally, equation (7) defining $n$ geometric constraints can be written in accordance with (12) from (10) and (11):

$$S_1 \subseteq ZT \Leftrightarrow \mathbf{t}_{N_i}.\mathbf{n}_i \le d^{\sup} \text{ and } \mathbf{t}_{N_i}.\mathbf{n}_i \le -d^{\inf}$$

$$\Leftrightarrow \mathbf{A}_1\mathbf{x} \le \mathbf{b}_1 \text{ and } \mathbf{A}_2\mathbf{x} \le \mathbf{b}_2 \Leftrightarrow \mathbf{Ax} \le \mathbf{b} \text{ with } \mathbf{A} = \begin{pmatrix} \mathbf{A}_1 \\ \mathbf{A}_2 \end{pmatrix} \text{ and } \mathbf{b} = \begin{pmatrix} \mathbf{b}_1 \\ \mathbf{b}_2 \end{pmatrix} \tag{12}$$

Using equation (12), it is possible to make equation (7) compliant as an $\mathcal{H}$ – description of a polyhedron $\mathcal{P}$ of $\mathbb{R}^6$ [12]. It defines $2n$ half-spaces $\overline{H_k}^-$ with $1 \le k \le 2n$. The intersection of the $2n$ half-spaces $\overline{H_k}^-$ defines a polyhedron $\mathcal{P}$ of $\mathbb{R}^6$.

$$\mathcal{P} = \left\{\mathbf{x} \in \mathbb{R}^6 : \mathbf{A}_1\mathbf{x} \le \mathbf{b}_1, \mathbf{A}_2\mathbf{x} \le \mathbf{b}_2\right\} = \left(\bigcap_{k_1=1}^{k_1=n} \overline{H_{k_1}}^-\right) \cap \left(\bigcap_{k_2=1}^{k_2=n} \overline{H_{k_2}}^-\right)$$

$$= \left\{\mathbf{x} \in \mathbb{R}^6 : \mathbf{Ax} \le \mathbf{b}\right\} = \bigcap_{k=1}^{k=2.n} \overline{H_k}^- \quad , k = k_1 + k_2 \tag{13}$$

Let us consider the example of Fig. 4(a) which shows the definition of an ISO specification for the location of a plane surface. This location implies that any manufactured surface $S_1$, a specific physical realization of the nominal square plane surface $S_0$ with dimension $b$, must be located within a tolerance zone defined by two parallel planes at $t_1$ distance apart: see Fig. 4(b). In the $(\mathbf{x},\mathbf{y},\mathbf{z})$ base, points A, B, C and D, vertices of the square of dimension $b$, have the following respective coordinates:

$$\begin{pmatrix} \frac{-b}{2} & \frac{b}{2} & 0 \end{pmatrix}, \begin{pmatrix} \frac{b}{2} & \frac{b}{2} & 0 \end{pmatrix}, \begin{pmatrix} \frac{b}{2} & \frac{-b}{2} & 0 \end{pmatrix} \text{ and } \begin{pmatrix} \frac{-b}{2} & \frac{-b}{2} & 0 \end{pmatrix}$$

The geometric constraints inherent in the location of $S_1$ are defined by eight half-spaces as in equation (13) [4]:

$$\mathcal{P} = \left\{\mathbf{x} = (r_x, r_y, r_z, t_{Nx}, t_{Ny}, t_{Nz})^T, \mathbf{Ax} \le \mathbf{b}\right\}$$

$$\text{with} : \mathbf{A} = \begin{pmatrix} b/2 & \mathrm{b}/2 & 0 & 0 & 0 & 1 \\ b/2 & -b/2 & 0 & 0 & 0 & -1 \\ b/2 & -b/2 & 0 & 0 & 0 & 1 \\ -b/2 & -b/2 & 0 & 0 & 0 & -1 \\ -b/2 & -b/2 & 0 & 0 & 0 & 1 \\ b/2 & \mathrm{b}/2 & 0 & 0 & 0 & -1 \\ -b/2 & \mathrm{b}/2 & 0 & 0 & 0 & 1 \\ b/2 & -b/2 & 0 & 0 & 0 & -1 \end{pmatrix} \text{ and } \mathbf{b} = \begin{pmatrix} \frac{t_1}{2} \end{pmatrix} \tag{14}$$

The intersection of the eight half-spaces, expressed at the center N of the square $(ABCD)$ with surface $S_0$ and generated by this location specification, generates a 3-dimension polytope in the affine space constructed on the $(r_x, r_y, t_{Nz})$ base, see Fig. 4(c).

## 3.2 Contact constraints

Like the geometric constraints described in 3.1, non-interferences by surfaces potentially in contact are characterized by contact constraints. Definition (7) can be specialized by writing:

$$\forall N_i \in E_c : d^{\inf} \leq \mathbf{t}_{N_i}.\mathbf{n}_i \leq d^{\sup} \quad (15)$$

With:

- $E_c$ : contact element (see Fig. 5) which may be along a surface, a line or defined by a point,
- $\mathbf{t}_{N_i}$ : translation displacement at point $N_i$ of surface 1 in relation to surface 2 between two parts potentially in contact,
- $\mathbf{n}_i$ : unit vector normal to a plane tangent to the two surfaces at point $N_i$, oriented towards the outer material in relation to surface 2,
- $d^{\inf}$ and $d^{\sup}$ are defined according to the type of contact: $d^{\inf} = -\frac{J}{2}$ and $d^{\sup} = +\frac{J}{2}$ in cases where the joints are in bilateral contact (generally the case with a cylindrical pair, ball-and-cylinder pair, spherical pair), where $J$ ( $J$ = dimension of hole - dimension of shaft) is clearance in the joint, the definition of $J$ is compliant with that given in the ISO standard [20].

A contact element is the element consisting of all possible points of contact between two surfaces. It derives from the intersection of two surfaces potentially in contact in the specific configuration where the two surfaces merge and there is no clearance.

For example, the contact element of a cylindrical pair joint is a cylinder. This cylinder is the intersection of two cylinders that are coaxial and have the same diameter. Fig. 5 shows the contact elements for planar pair and cylindrical pair joints (surface contact elements), cylinder- and-plane pair and ball-and-cylinder pair joints (line contact element), and ball-and-plane pair joint (ponctual contact element). This non-exhaustive list gives the main types of joint used in tolerance analysis.

The definition of a contact polyhedron $\mathcal{P}_1$ is similar to that in equation (13).

In cases of unilateral contact (usually ball-and-plane pair joint, cylinder-and-plane pair or planar pair joints), we have the specific case where:

- $d^{\inf} = 0$,
- $d^{\sup}$ is not defined.

The result is that only equations $d^{\inf} \leq \mathbf{t}_{N_i}.\mathbf{n}_i$ defined in (12) are generated by the contact constraints. The polyhedron is defined as follows:

$$\mathcal{P}_2 = \left\{ \mathbf{x} \in \mathbb{R}^6 : \mathbf{A}_2 \mathbf{x} \leq \mathbf{b}_2 \right\} = \bigcap_{k_2=1}^{k_2=n} \overline{H_{k_2}^{-}} \quad (16)$$

## 3.3 Geometric polytope, contact polytope

Generally, the relative position of any two surfaces of a mechanical system derives from the sum and intersection of $\mathbb{R}^6$ polyhedra. Due to the algorithmic complexity of Minkowski sums, we chose to manipulate only geometric polytopes and contact polytopes, in other words bounded sets. A geometric polytope is defined as a bounded geometric polyhedron. Similarly, a contact polytope is a bounded contact polyhedron. In the next two sections we show how geometric and contact polyhedra are made compliant with geometric and contact polytopes respectively.

### 3.3.1 Geometric polytope

Geometric polyhedra defined previously in (13) are not generally bounded as their surfaces usually have degrees of invariance [21]. The degree of invariance of a surface will hereafter be written $d_{inv}$. These polyhedra are therefore not $\mathbb{R}^6$ polytopes.

In this article, we propose to put boundaries on these polyhedra by the addition of $m$ $\overline{H_c}^-$ half-spaces, called *cap* half-spaces, to all the half-spaces resulting from the geometric constraints. The degree of invariance of a surface depends on the class of surface [21]; it defines a theoretically unlimited displacement which leaves the toleranced surface in the Euclidean space invariant overall. A displacement may have two boundaries, each one characterizing a cap half-space. Thus, two cap half-spaces per degree of invariance are added to the half-spaces defining the polyhedron (see Table 1). Generally in $\mathbb{R}^6$, the geometric polytope derived from this addition of $m = 2.d_{inv}$ cap half-spaces can be defined by the expression:

$$\mathcal{P} = \left( \bigcap_{k=1}^{k=2.n} \overline{H_k}^- \right) \cap \left( \bigcap_{c=1}^{c=2.d_{inv}} \overline{H_c}^- \right) \tag{17}$$

Where: $d_{inv}$ is the degree of invariance of the surface.

The cap half-spaces are additional half-spaces so that a $\mathbb{R}^6$ polyhedron can be made compliant with a 6-polytope while still retaining the geometry and the topology of the polyhedron, i.e. the polyhedron faces. This polytope defines:

- displacements of the surface inside a tolerance zone,
- displacements limited by the cap half-spaces that leave the surface invariant overall.

Half-spaces identified in equation (17) can be written:

$$\begin{aligned} \overline{H_k}^- &= \mathbf{a}_k{}^T \mathbf{x} \le b_k ,\, b_k = d^{\sup} \text{ or } b_k = -d^{\inf} \\ \overline{H_c}^- &= \mathbf{a}_c{}^T \mathbf{x} \le C ,\, C \in ]0,+\infty[ \end{aligned} \tag{18}$$

The second members of the half-spaces correspond to the distances of their respective hyper-planes from the origin of the displacements, see Fig. 6.

The second member of a half-space generated by a geometric constraint depends on $d^{\sup}$ or $d^{\inf}$ that is to say on the dimension of the tolerance zone $t$ and the position of the tolerance zone in relation to the nominal surface, see Fig. 6.

The second member of a cap half-space $\overline{H_c}^-$, identified as $C$, is a strictly positive value.

The cap half-spaces are introduced for the sole purpose of facilitating the calculation of the sums and intersections of the polyhedra (unbounded sets) by transforming them into polytopes (bounded sets). The value of $C$ must be large enough to distinguish bounded displacements from unbounded ones, while not affecting the final result. The choice of the value of $C$ will be considered in further detail later in the article.
The $\mathbf{a}_c$ coefficients depend on the classes of the surfaces and the displacements leaving the surfaces invariant overall.

A $(\mathbf{u},\mathbf{v},\mathbf{w})$ base is associated to each surface according to the displacement axes, leaving it invariant overall. The cap half-spaces enabling them to be limited are defined on the basis of these axes: see Fig. 7.

Let us return to the example in Fig. 4: this shows a plane surface where $d_{inv}=3$. In $\mathbb{R}^6$, $2.d_{inv}=6$ cap half-spaces at least are required to set boundaries to the intersection of the half-spaces resulting from geometric constraints (13). In a $(\mathbf{u},\mathbf{v},\mathbf{w})$ base linked to the plane surface such that $\mathbf{w}$ is normal to the surface, displacements leaving the surface invariant overall are limited as follows, with $C>0$ and N the point of definition of half-spaces $\overline{H_k^-}$ :

$$\begin{cases} \mathbf{r}.\mathbf{w}\le C & -\mathbf{r}.\mathbf{w}\le C \\ \mathbf{t}_N.\mathbf{u}\le C \;, & -\mathbf{t}_N.\mathbf{u}\le C \\ \mathbf{t}_N.\mathbf{v}\le C & -\mathbf{t}_N.\mathbf{v}\le C \end{cases} \tag{19}$$

Equations at (19) can be made compliant wih 6 cap half-spaces of $\mathbb{R}^6$ in the same way as those at (18). We now define the following projections for any $(\mathbf{x},\mathbf{y},\mathbf{z})$ base:

$$\mathbf{u}\begin{pmatrix} u_x \\ u_y \\ u_z \end{pmatrix}, \mathbf{v}\begin{pmatrix} v_x \\ v_y \\ v_z \end{pmatrix} \text{ and } \mathbf{w}\begin{pmatrix} w_x \\ w_y \\ w_z \end{pmatrix} \tag{20}$$

the 6 cap half-spaces are then defined such that:

$$\begin{cases} r_x.w_x + r_y.w_y + r_z.w_z \le C \\ -r_x.w_x - r_y.w_y - r_z.w_z \le C \\ t_{Nx}.\mathrm{u}_x + t_{Ny}.u_y + t_{Nz}.u_z \le C \\ -t_{Nx}.\mathrm{u}_x - t_{Ny}.u_y - t_{Nz}.u_z \le C \\ t_{Nx}.\mathrm{v}_x + t_{Ny}.\mathrm{v}_y + t_{Nz}.\mathrm{v}_z \le C \\ -t_{Nx}.\mathrm{v}_x - t_{Ny}.\mathrm{v}_y - t_{Nz}.\mathrm{v}_z \le C \end{cases} \tag{21}$$

Fig. 8 shows the limits of surface displacements in the $(r_x,r_y,t_x)$ base where displacement according to $t_{Nx}$ is unbounded and the corresponding cap half-spaces.

Cap half-spaces are similarly defined for classes of surface: spherical, cylindrical, revolution or prismatic.

### 3.3.2 Contact polytope

The contact polyhedra defined above are generally unbounded as the joints have degrees of freedom $d_{mob}$. Just as the limits of surface displacement in a tolerance zone can be defined, displacement limits for two surfaces potentially in contact can be modeled in the case of bilateral contacts. Cap half-spaces limit the displacements inherent in the degrees of freedom of the joints.
The number of cap half-spaces in this case depends on the type of joint (see Table 2).
Each degree of freedom defines a theoretically unbounded displacement which is limited by two cap half-spaces. Finally, $2.d_{mob}$ cap half-spaces are added to the half-spaces generated by contact constraints, by analogy with the $2.d_{inv}$ cap half-spaces added to the half-spaces generated by geometric constraints according to (17).

In the specific case of joints with unilateral contact, each contact constraint limits displacement in a single direction (16). Contact constraints generate only $n$ half-spaces of $\mathbb{R}^6$ instead of $2n$. In order to set boundaries on the intersection of these $n$ half-spaces, $n$ other constraints are added so as to suppress the unilaterality of the contact. These $n$ constraints form $n$ extra cap half-spaces in addition to the $m$ cap half-spaces limiting displacements resulting from the degrees of freedom according to equation (22) where $d_{mob}$ corresponds to the degree of freedom of the joint:

$$\mathcal{P}_2 = \left( \bigcap_{k_2=1}^{k_2=n} \overline{H_{k_2}}^- \right) \cap \left( \bigcap_{c=1}^{c=m+n} \overline{H_c}^- \right) \quad \text{with } m = 2.d_{mob} \tag{22}$$

In this case, the polyhedron (16) is defined solely by matrix $\mathbf{A}_2$. To suppress the unilaterality of the contact, we propose to add $n$ constraints identified by matrix $\mathbf{A}_1$ (dimension $n$) opposite matrix $\mathbf{A}_2$. Next, the $m$ cap half-spaces are added, according to the degrees of freedom.

Just as for the cap half-spaces for the geometric polytopes, the value of $C$ for the second members should be selected with care so as not to affect the final result.

## 3.4 Synthesis

An operand polytope will then be defined as follows:

$$\mathcal{P} = \left( \bigcap_{k=1}^{k=n_k} \overline{H_k}^- \right) \cap \left( \bigcap_{c=1}^{c=n_c} \overline{H_c}^- \right) \quad \text{with} \quad n_k + n_c = 2n + m \tag{23}$$

Where $\overline{H_k}^-$ is the set of $n_k$ half-spaces resulting from the geometric or contact constraints and $\overline{H_c}^-$ is the set of $n_c$ cap half-spaces. The intersection of the $n_k$ half-spaces defines a polyhedron.

We now propose the following definitions:

**Definition 1**: a $j$-face of a polytope is described as a non-cap $j$-face if and only if it is included in a $j$-face of the corresponding polyhedron.

**Definition 2**: any $j$-face that is not a non-cap $j$-face is cap.

For geometric or contact polytopes, the non-cap facets are thus included in the hyper-planes $H_k$, boundaries of the half-spaces $\overline{H_k}^-$ resulting from the geometric or contact constraints.

Fig. 9 shows a synthesis of the tolerance analysis procedure by polytopes. From the tolerance zones on the one hand and the joints on the other, the geometric and contact constraints can be written respectively.
These constraints usually define unbounded sets, or polyhedra.
Cap half-spaces are added to the polyhedra in order to set boundaries and transform them into polytopes. The sums and intersections of these polytopes can then be calculated to simulate geometric variations according to the structure of the mechanism by a resulting polytope. From this, the boundaries of the functional defects of the mechanism can be identified and we are thus able to determine whether the limits set out in the specifications are respected.

The next paragraph describes the methods used to differentiate among all the facets of a polytope resulting from a sum or an intersection, those facets that are generated by cap half-spaces from the others.

# 4 Identification of caps in the double description of a Minkowski sum of polytopes

## 4.1 Problem

Let us define polytope $\mathcal{P}_3$ resulting from the Minkowski sum of two polytopes $\mathcal{P}_1$ and $\mathcal{P}_2$ of $\mathbb{R}^n$ as follows:

$$\mathcal{P}_3 = \mathcal{P}_1 \oplus \mathcal{P}_2 = \{p + q \mid p \in \mathcal{P}_1, q \in \mathcal{P}_2\} \quad (24)$$

We shall try to define the faces of polytope $\mathcal{P}_3$ which are independent from the cap half-spaces of operands $\mathcal{P}_1$ and $\mathcal{P}_2$ by manipulating the polytopes using a double description (i.e. by their $\mathcal{V}$ – description and their $\mathcal{H}$ – description).

Fukuda has shown that every face of the sum $\mathcal{P}_3$ of two $n$-polytopes $\mathcal{P}_1$ and $\mathcal{P}_2$ can be decomposed in only one way into the sum of two faces of operands $\mathcal{P}_1$ and $\mathcal{P}_2$ [18].

Let us propose the following:
**Property 1:** the faces of $\mathcal{P}_3$ resulting from the sum of two non-cap faces of $\mathcal{P}_1$ and $\mathcal{P}_2$ are the non-cap faces of $\mathcal{P}_3$. All the other faces of $\mathcal{P}_3$ are called cap faces.

Fig. 10 shows a classic example of the Minkowski sum of two polytopes $\mathcal{P}_1$ and $\mathcal{P}_2$ in $\mathbb{R}^2$ in geometric tolerancing. Polytope $\mathcal{P}_1$ is made up of four non-cap facets $e_{11}, e_{12}$, $e_{13}$ and $e_{14}$ (solid line in Fig. 10). Polytope $\mathcal{P}_2$ is made up of two non-cap facets $e_{21}$ and $e_{23}$ (solid line in Fig. 10) and two cap facets $e_{22}$ and $e_{24}$ (dotted line in Fig. 10).

All the vertices of $\mathcal{P}_1$ are non-cap whereas all the vertices of $\mathcal{P}_2$ are cap. In fact, polytope $\mathcal{P}_1$ is identical to the corresponding polyhedron and the vertices of $\mathcal{P}_1$ are therefore vertices of the polyhedron, and hence are non-cap, according to definition 1 given in §3.4. The vertices of $\mathcal{P}_2$ do not exist on the corresponding polyhedron: they are therefore cap, according to definitions 1 and 2 given in §3.4.

We will now consider two examples of sums of faces (edges or vertices) in this example. The first is the sum of the two non-cap edges $e_{12}$ and $e_{23}$. The result is edge $e_{33}$ of $\mathcal{P}_3$ where: $e_{33} = e_{12} \oplus e_{23}$.
$e_{33}$ is a non-cap edge according to property 1, see Fig. 10.

The second example is the Minkowski sum of a non-cap edge $e_{11}$ of $\mathcal{P}_1$ and a cap vertex $v_{2a}$ of $\mathcal{P}_1$ and $\mathcal{P}_2$ respectively. The result is a cap edge $e_{31}$ of $\mathcal{P}_3$ where $e_{31} = e_{11} \oplus v_{2a}$.
$e_{31}$ is a cap edge according to property 1, see Fig. 10.

The second example shows that a non-cap facet of an operand polytope can be translated into a cap facet in the polytope sum. In other words, a cap facet in the first operand can modify the attribute (from non-cap to cap) of a facet in the polytope sum, as a result of the translation of a facet from the second operand.
In $\mathbb{R}^6$, the algorithmic application of the property posited by Fukuda in [18] requires all the d-faces $(0 \le d \le 5)$ of operands $\mathcal{P}_1$ and $\mathcal{P}_2$ to be determined in order to identify all the cap facets in $\mathcal{P}_3$. This is very complex to implement with dimensions greater than 3 and is prohibitive in terms of algorithmic complexity and hence in calculation time.

For this reason we decided to manipulate the polytopes by their $\mathcal{V}-$description and their $\mathcal{H}-$description, i.e. with a double description [4], [14]. This guarantees the traceability of the specifications in operations used in geometric tolerancing [14], however it is then difficult to distinguish cap faces and non-cap faces since property 1 cannot be applied.

## 4.2 Algorithm to calculate the Minkowski sum of polytopes

The method proposed here uses the normal fans of operand polytopes and the normal fans of associated operand polyhedra. To show how cap faces can be identified through normal fans, we first describe the basis of the algorithm used to determine the Minkowski sum of polytopes.
The topology of polytope $\mathcal{P}_3 = \mathcal{P}_1 + \mathcal{P}_2$ is defined by its normal fan $N(\mathcal{P}_3)$ resulting from a common refinement of the normal fans of operand polytopes [17]:

$$N(\mathcal{P}_3) = N(\mathcal{P}_1) \wedge N(\mathcal{P}_2) \tag{25}$$

Determining the common refinement of two fans $N(\mathcal{P}_1)$ and $N(\mathcal{P}_2)$ involves calculating the union of all the intersections of the polyhedric cones of fans $N(\mathcal{P}_1)$ and $N(\mathcal{P}_2)$ taken two by two [12].
Polytope $\mathcal{P}_1$ is characterized by its vertices $v_{1i}$ where $v_{1i}$ is the i$^{th}$ vertex of $\mathcal{P}_1$.
Similarly:

- $\mathcal{P}_2$ is characterized by its vertices $v_{2j}$ where $v_{2j}$ is the j$^{th}$ vertex of $\mathcal{P}_2$,
- $\mathcal{P}_3$ is characterized by its vertices $v_{3k}$ where $v_{3k}$ is the k$^{th}$ vertex of $\mathcal{P}_3$.

Let $\text{DualCone}(v_{1i})$, $\text{DualCone}(v_{2j})$ and $\text{DualCone}(v_{3k})$ be the dual cones associated with vertices $v_{1i}$, $v_{2j}$ and $v_{3k}$ respectively.

We have shown in [14] the property (26) in $\mathbb{R}^n$:

$$\begin{aligned}&\dim\left(\text{DualCone}(v_{1i})\cap\text{DualCone}(v_{2j})\right)=n\Rightarrow\\&\quad v_{1i}+v_{2j}=v_{3k}\\&\quad \text{DualCone}(v_{1i})\cap\text{DualCone}(v_{2j})=\text{DualCone}(v_{3k})\\&\quad \text{where the edges of DualCone}(v_{3k})\text{ are normal to the facets of }\mathcal{P}_3\text{ that converge at }v_{3k}\end{aligned} \tag{26}$$

Equations (24) and (25) can be generalized to polyhedra. If $\mathcal{P}_1$ and $\mathcal{P}_2$ are polyhedra, $\mathcal{P}_3$ is also a polyhedron; the normal fans $N^R(\mathcal{P}_1)$, $N^R(\mathcal{P}_2)$ and $N^R(\mathcal{P}_3)$ are not complete, i.e. they do not partition $\mathbb{R}^n$. The normal fan of a polyhedron is a regular Gröbner fan [22], [23].
In general, $N(\mathcal{P}_1)$ designates the normal fan of polytope $\mathcal{P}_1$ and $N^R(\mathcal{P}_1)$ designates the normal fan of the polyhedron containing $\mathcal{P}_1$.

## 4.3 Identification of cap facets in the $\mathcal{H}$ – description of a polytope sum

To identify cap and non-cap facets, we define the following property:

***Property 2:*** *The normals of the non-cap facets of $\mathcal{P}_3$ are the result of the common refinement of the normal fans $N^R(\mathcal{P}_1)$ and $N^R(\mathcal{P}_2)$ of the corresponding operand polyhedra.*

Fig. 11(a) shows the normal fan $N(\mathcal{P}_3)$ of polytope $\mathcal{P}_3$, resulting from the common refinement of fans $N(\mathcal{P}_1)$ and $N(\mathcal{P}_2)$. The edges of the dual cones that make up $N(\mathcal{P}_3)$ are the normal of the facets of $\mathcal{P}_3$: see Fig. 10 and Fig. 11(a).

**Note**: the normal fans shown are normalized normal fans, in other words, they are intersected with the unit sphere [24].

Fig. 11(b) shows the normal fan $N^R(\mathcal{P}_3)$, common refinement of the normal fans $N^R(\mathcal{P}_1)$ and $N^R(\mathcal{P}_2)$ of the polyhedra associated with $\mathcal{P}_1$ and $\mathcal{P}_2$ respectively. The polyhedron associated with $\mathcal{P}_1$ results from the intersection of the half-spaces defining $\mathcal{P}_1$ without its cap half-spaces. Likewise for the polyhedron associated with $\mathcal{P}_2$.
In the example in Fig. 10, the polyhedron associated with $\mathcal{P}_1$ and polytope $\mathcal{P}_1$ are identical, since $\mathcal{P}_1$ has no cap half-spaces. The same is true for their normal fans. The normal fan $N^R(\mathcal{P}_2)$ of the polyhedron associated with $\mathcal{P}_2$ is reduced to two rays (two 1 dimension cones). Similarly for the common refinement $N^R(\mathcal{P}_1)\wedge N^R(\mathcal{P}_2)$.

The edges of the normal fan $N^R(\mathcal{P}_3)$ are normals of the facets of the polyhedron Minkowski sum of the polyhedra associated with $\mathcal{P}_1$ and $\mathcal{P}_2$ respectively.
From this we deduce that the normals of the non-cap facets of $\mathcal{P}_3$ correspond to the edges of $N(\mathcal{P}_3)$ which are common to the edges of $N^R(\mathcal{P}_3)$, i.e. edges $e_{33}$ and $e_{36}$ , see Fig. 10 and Fig. 11(b).

## 4.4 Identifying cap vertices in $\mathcal{V}$ – description of a sum polytope

The non-cap vertices of $\mathcal{P}_3$ are easy to identify.
From equation (26) and property 1 taken from the work of Fukuda [18], we can deduce equations (27) and (28):

if $v_{1i}$ and $v_{2j}$ are non-cap vertices,
$$\dim\left(\text{DualCone}\left(v_{1i}\right)\cap\text{DualCone}\left(v_{2j}\right)\right)=n\Rightarrow v_{1i}+v_{2j}=v_{3k} \tag{27}$$
where $v_{3k}$ is a non-cap vertex

if $v_{1i}$ or $v_{2j}$ is a cap vertex,
$$\dim\left(\text{DualCone}\left(v_{1i}\right)\cap\text{DualCone}\left(v_{2j}\right)\right)=n\Rightarrow v_{1i}+v_{2j}=v_{3k} \tag{28}$$
where $v_{3k}$ is a cap vertex

# 5 Identification of caps in the double description of a polytope intersection

## 5.1 Problem

The intersection of two operand polytopes $\mathcal{P}_1$ and $\mathcal{P}_2$ is defined as the intersection of the set of half-spaces from the $\mathcal{H}$-descriptions of the two operands:
$$\mathcal{P}_1\cap\mathcal{P}_2=\left(\bigcap_{u=1}^{u=n_1}\overline{H_u}^{-}\right)\cap\left(\bigcap_{v=1}^{v=n_1}\overline{H_v}^{-}\right)\quad\text{with}\quad\mathcal{P}_1=\bigcap_{u=1}^{u=n_1}\overline{H_u}^{-}\quad\text{and}\quad\mathcal{P}_2=\bigcap_{v=1}^{v=n_2}\overline{H_v}^{-} \tag{29}$$

According to definition (23), we have:
$$\mathcal{P}_i=\left(\bigcap_{k_i=1}^{k_i=n_{k_i}}\overline{H_{k_i}}^{-}\right)\cap\left(\bigcap_{c_i=1}^{c_i=n_{c_i}}\overline{H_{c_i}}^{-}\right)\quad\text{with}\quad n_{k_i}+n_{c_i}=n_i \tag{30}$$
Where: $n_{k_i}$ is the number of non-cap half-spaces denoted $\overline{H_{k_i}}^{-}$ and $n_{c_i}$ is the number of cap half-spaces denoted $\overline{H_{c_i}}^{-}$ .
In geometric tolerancing, we need to determine the intersection of the displacements defined by the geometric and/or contact constraints. This intersection derives from the intersection of the non-cap half-spaces of operands $\mathcal{P}_1$ and $\mathcal{P}_2$, i.e. polyhedra $\mathcal{P}_1^{(R)}$ and $\mathcal{P}_2^{(R)}$ containing $\mathcal{P}_1$ and $\mathcal{P}_2$ respectively, see Fig. 12(a). In order to make this intersection compliant with a polytope $\mathcal{P}_3$ (i.e. a bounded set), a set of non-cap half-spaces must be added, see Fig. 12(b). Polytope $\mathcal{P}_3$ is defined by equation (31) in accordance with (23):

$$\mathcal{P}_3 = \left( \bigcap_{k_1=1}^{k_1=n_{k_1}} \overline{H_{k_1}}^- \right) \cap \left( \bigcap_{k_2=1}^{k_2=n_{k_2}} \overline{H_{k_2}}^- \right) \cap \left( \bigcap_{c_3=1}^{c_3=n_{c_3}} \overline{H_{c_3}}^- \right) \tag{31}$$

where $\overline{H_{c_3}}^-$ denotes a cap half-space of $\mathcal{P}_3$.

It is no easy matter to determine cap half-spaces $\overline{H_{c_3}}^-$. The topology of polytope $\mathcal{P}_1 \cap \mathcal{P}_2$ depends on the choice of second member $C$ of the cap half-spaces of the operands $\mathcal{P}_1$ and $\mathcal{P}_2$. Thus as a general rule: $\mathcal{P}_3 \neq \mathcal{P}_1 \cap \mathcal{P}_2$, see Fig. 12(c).

In the next section we propose a method to truncate a hyper-parallelepiped, to determine displacements limited by the non-cap half-spaces resulting from the intersection of two operand polytopes.

## 5.2 Determining the cap half-spaces of the intersection of two polytopes

One method to determine $\mathcal{P}_3$ consists of determining the intersection between:

- a right hyper-parallelepiped circumscribed at the union of the vertices of $\mathcal{P}_1$ and $\mathcal{P}_2$,
- the non-cap half-spaces of operands $\mathcal{P}_1$ and $\mathcal{P}_2$.

The facets of polytope $\mathcal{P}_3$ where the hyper-planes contain the facets of the right hyper-parallelepiped are the cap facets of $\mathcal{P}_3$, which can be used to define $\left( \bigcap_{c_3=1}^{c_3=n_{c_3}} \overline{H_{c_3}}^- \right)$ introduced into equation (31). All the other facets are non-cap. In addition, all the vertices of $\mathcal{P}_3$ located on a hyper-plane containing one of the facets of the initial right hyper-parallelepiped are cap vertices. All the other vertices are non-cap.

Fig. 13(a) shows the determination of a circumscribed right hyper-parallelepiped (a rectangle in this example) at the union of the vertices of $\mathcal{P}_1$ and $\mathcal{P}_2$.
The circumscribed right hyper-parallelepiped must not be a minimal volume. If this were the case, then this may result in a hyper-parallelepiped of which one facet might be confused with a facet of $\mathcal{P}_1$ or $\mathcal{P}_2$. It would then not be possible to detect whether such a facet is cap or non-cap.

To avoid this, a right circumscribed hyper-parallelepiped is determined at an interval $d$ from the minimal volume circumscribed right hyper-parallelepiped, where $d > 0$: Fig. 13(a).
Fig. 13(b) shows polytope $\mathcal{P}_3$ resulting from the intersection between a circumscribed right hyper-parallelepiped with vertices $\mathcal{P}_1$ and $\mathcal{P}_2$ and the non-cap half-spaces of operands $\mathcal{P}_1$ and $\mathcal{P}_2$.

# 6 Application to a mechanical system

We shall now apply the methods for identifying cap half-spaces in the sums and intersections of polytopes using a simulation to ensure that a functional condition is fulfilled in a simple mechanical system, illustrated in Fig. 14. This system is made up of three parts: support 1, bearing 2 and shaft 3. No degree of freedom is allowed between support 1 and bearing 2. Two degrees of freedom are allowed between the shaft 3 and the bearing 2: rotation and translation along $\mathbf{x}$.
A functional condition FC is defined across the distance $d$ between point M of the shaft 3 and the support 1:

$$d\min \le d \le d\max \tag{32}$$

Considering $d_0$ the nominal value of $d$, $dev$, $dev_{\inf}$ and $dev_{\sup}$ such that:

$$dev = d - d_0$$
$$dev_{\inf} = d\min - d_0$$
$$dev_{\sup} = d\max - d_0$$

The equation can still be written:

$$dev_{\inf} \le dev \le dev_{\sup} \tag{33}$$

The variable $dev$ is the deviation in translation along $\mathbf{y}$ of point M on shaft 3 in relation to support 1, with a lower bound of $dev_{\inf}$ and an upper bound of $dev_{\sup}$.

The purpose of this application is to simulate the compliance of the mechanical system in terms of the functional condition FC, given the geometric variations assigned to parts 1, 2 and 3.

This example will be considered in the normal plane $\mathbf{z}$ containing point M. Thus each displacement will be defined by three parameters: rotation along $\mathbf{z}$, translation along $\mathbf{x}$ and translation along $\mathbf{y}$. The polytopes manipulated will be 3 dimension and can be represented graphically.

## 6.1 Description of the mechanical system

Parts 1, 2 and 3 are made up solely of plane and cylindrical surfaces.
We write the surface j of part i as i,j.
Surfaces 2,4 and 3,4 are cylindrical; surfaces 1,1; 1,2; 1,3; 1,4; 2,1; 2,2 and 2,3 are plane surfaces, see Fig. 15.

The topological structure of the mechanical system is set out in a graph [25], see Fig. 15.

For each surface i,j there is a corresponding vertex i,j on the graph. Vertices 1,0; 2,0 and 3,0 correspond to nominal models of parts 1, 2 and 3 respectively: see Fig. 15.

An edge connected to a vertex designated i,0 (i= 1, 2 or 3) corresponds to position deviations of a surface i,j in relation to its nominal position.

A geometric polytope is associated to each edge connected to a vertex designated i,0 (i= 1, 2 or 3). This polytope characterizes the variations in position of a surface i,j in relation to its nominal surface inside a tolerance zone of dimension $t_{i,j}$.
For a plane surface i,j, $t_{i,j}$ is the distance between two parallel planes arranged symmetrically in relation to a nominal plane as shown in Fig. 4.
For a cylindrical surface, $t_{i,j}$ is the diameter of a coaxial cylinder with the axis of the corresponding nominal cylinder.

Parts 1, 2 and 3 are positioned according to four contact specifications [26] so that:

- contact between surfaces 1,1 and 2,1 is "unilateral", planar pair type, contact element normal plane $\mathbf{y}$, nature of contact is sliding (i.e. clearance $J_1$ between 1,1 and 2,1 is null).
- two contacts between surfaces 1,2 / 2,2 and 1,3 / 2,3 respectively are unilateral, cylinder-and-plane pair type, contact element is a straight line parallel to $\mathbf{z}$ contained within a normal plane $\mathbf{x}$, nature of contact is floating.
- contact between 3,4 and 2,4 is bilateral, cylindrical pair type, contact element is a cylinder with director vector $\mathbf{x}$, nature of contact is floating (clearance $J_4$ between 2,4 and 3,4 is such that: $J_4 = D_{2,4} - D_{3,4}$ with $J_4 > 0$).

An edge between two surfaces of two distinct parts corresponds to the position deviations between two surfaces potentially in contact.
A contact polytope is associated to each edge connected between two surfaces of two distinct parts. The four contact polytopes of the system studied here derive from the four contact specifications formulated above; they will be described in more detail in the following section.

Finally, an additional edge, labeled FC defines the functional condition between surfaces 3,4 and 1,4. A "functional polyhedron" is associated to this edge, derived from the functional constraints defined in equation(33).

## 6.2 Definition of operand polytopes

The different points (I, J, L, M, N …) marked in Fig. 16 define the geometric constraints, contact constraints and cap half-spaces needed to characterize operand polytopes.

Table 3 summarizes the definition of geometric polytopes associated to the respective surfaces identified in Fig. 15, and according to the procedure described in §3.3.1.
For geometric polytopes associated to surfaces 2,2 and 2,3 respectively, we consider $L_4 = L_5$, see Fig. 16.

We designate as $\mathcal{P}_{i,j/\mathrm{i},0}$ the geometric polytope characterizing position variations between surface i,j and its nominal model i,0.

Table 3. Definitions of geometric polytopes.
summarizes the definition of contact polytopes associated respectively to the four contact specifications of the system studied, according to the procedure described in §3.3.2.
We designate as $\mathcal{P}_{i,j/u,v}$ the contact polytope characterizing the position variations between surface i,j and surface u,v.

## 6.3 Simulation of mechanical system compliance with FC

Because of the topological structure of the graph it is possible to define the operations required between operand polytopes to determine polytope $\mathcal{P}_{3,4/1,4}$ characterizing variations in position between surfaces 3,4 and 1,4 [4], [16].

Polytope $\mathcal{P}_{3,4/1,4}$ is defined by:

$$\mathcal{P}_{3,4/1,4} = \mathcal{P}_{3,4/2,4} \oplus \mathcal{P}_{2,4/2,0} \oplus \begin{bmatrix} \left(\mathcal{P}_{2,0/2,1} \oplus \mathcal{P}_{2,1/1,1} \oplus \mathcal{P}_{1,1/1,0}\right) \cap \left(\mathcal{P}_{2,0/2,2} \oplus \mathcal{P}_{2,2/1,2} \oplus \mathcal{P}_{1,2/1,0}\right) \\ \cap \left(\mathcal{P}_{2,0/2,3} \oplus \mathcal{P}_{2,3/1,3} \oplus \mathcal{P}_{1,3/1,0}\right) \end{bmatrix} \oplus \mathcal{P}_{1,0/1,4} \quad (34)$$

For easier writing and in accordance with the method described in §5, the operator $\cap$ designates an operation to determine the intersection between:

- a circumscribed right hyper-parallelepiped at the union of the operand vertices,
- the non-cap half-spaces of the operand polytopes.

The compliance of the mechanical system is validated by including the calculated polytope $\mathcal{P}_{3,4/1,4}$ in the functional polyhedron $\mathcal{P}^{f}_{3,4/1,4}$ according to the resultant of the FC condition defined in:

$$\mathcal{P}_{3,4/1,4} \subseteq \mathcal{P}^{f}_{3,4/1,4} \quad (35)$$

The equation can be written as:

$$\begin{aligned} &\mathcal{P}_{3,4/1,4} = \mathcal{P}_{3,4/2,4} \oplus \mathcal{P}_{2,4/2,0} \oplus \mathcal{P}_{1,0/1,1} \oplus \left(\mathcal{P}_1 \cap \mathcal{P}_2 \cap \mathcal{P}_3\right) \\ &\text{with :} \\ &\mathcal{P}_1 = \mathcal{P}_{2,0/2,1} \oplus \mathcal{P}_{2,1/1,1} \oplus \mathcal{P}_{1,1/1,0} \\ &\mathcal{P}_2 = \mathcal{P}_{2,0/2,2} \oplus \mathcal{P}_{2,2/1,2} \oplus \mathcal{P}_{1,2/1,0} \\ &\mathcal{P}_3 = \mathcal{P}_{2,0/2,3} \oplus \mathcal{P}_{2,3/1,3} \oplus \mathcal{P}_{1,3/1,0} \end{aligned} \quad (36)$$

Firstly, polytopes $\mathcal{P}_1$, $\mathcal{P}_2$ and $\mathcal{P}_3$ are determined by applying the method described in §4. Polytopes $\mathcal{P}_2 \cap \mathcal{P}_3$ and $\mathcal{P}_1 \cap \mathcal{P}_2 \cap \mathcal{P}_3$ are then determined using the method described in §5. Lastly, polytope $\mathcal{P}_{3,4/1,4}$ is deduced from $\mathcal{P}_1 \cap \mathcal{P}_2 \cap \mathcal{P}_3$ and other operands by applying the method in §4.
To simplify the graphs, the following points have been considered:

- the dimensions of the tolerance zones $t_{i,j}$ are all equal one with another, with $t_{i,j} > 0$,
- clearance $J_4$ is such that: $J_4 = t_{i,j}$.

For the graphics:

- facets shown in shaded mode are non-cap facets,
- facets shown in wireframe mode are cap facets.

### 6.3.1 Determination of $\mathcal{P}_1$, $\mathcal{P}_2$ and $\mathcal{P}_3$

Polytopes $\mathcal{P}_{1,1/1,0}$ and $\mathcal{P}_{2,1/2,0}$ are identical and their center of symmetry is the origin of the displacements; only facets orthogonal to direction $t_{M-x}$ are cap facets: see Table 3. Polytope $\mathcal{P}_{2,1/1,1}$ may be considered as a polytope reduced to a single vertex corresponding to the origin of the displacements, given the contact constraints defined in Table 3. Definitions of geometric polytopes.
. Fig. 17(a) shows the determining of polytope $\mathcal{P}_1$ ($\mathcal{P}_1 = \mathcal{P}_{2,0/2,1} \oplus \mathcal{P}_{2,1/1,1} \oplus \mathcal{P}_{1,1/1,0}$) in $\mathbb{R}^3$ where $\mathcal{P}_{2,1/1,1}$, a neutral element of the Minkowski sum, is not shown. Fig. 17(b) shows a projection of polytope $\mathcal{P}_1$ in $\mathbb{R}^2$, in the plane $\left(r_z, t_{My}\right)$. In this example, the topology of polytope $\mathcal{P}_1$ is identical to the topologies of $\mathcal{P}_{1,1/1,0}$ and $\mathcal{P}_{2,1/2,0}$ given that the Minkowski sum between two isometric polytopes resembles a homothety coefficient of 2 [12]. Only the facets of $\mathcal{P}_1$ orthogonal to direction $t_{Mx}$ are cap facets.

Polytopes $\mathcal{P}_{1,2/1,0}$ and $\mathcal{P}_{2,2/2,0}$ are identical and their center of symmetry is the origin of the displacements, see Table 3. The non-cap facets of $\mathcal{P}_{1,2/1,0}$ and $\mathcal{P}_{2,2/2,0}$ are those which are orthogonal to direction $t_{My}$.

Polytope $\mathcal{P}_{2,2/1,2}$ is not centered at the origin of the displacements and has only one non-cap facet corresponding to its only contact constraint, see Table 3. Definitions of geometric polytopes.

.

By applying the method developed in §4, we can deduce polytope $\mathcal{P}_2$ (with $\mathcal{P}_2 = \mathcal{P}_{2,0/2,2} \oplus \mathcal{P}_{2,2/1,2} \oplus \mathcal{P}_{1,2/1,0}$) shown in $\mathbb{R}^3$ in Fig. 18(a).

Fig. 18(b) shows a projection of polytope $\mathcal{P}_2$ in $\mathbb{R}^2$, in the plane $\left(r_z, t_{My}\right)$.

Polytope $\mathcal{P}_2$ has only one non-cap facet resulting from the only non-cap facet of contact polytope $\mathcal{P}_{2,2/1,2}$.

Similarly, Fig. 19(a) shows the determination of $\mathcal{P}_3$ ($\mathcal{P}_3 = \mathcal{P}_{2,0/2,3} \oplus \mathcal{P}_{2,3/1,3} \oplus \mathcal{P}_{1,3/1,0}$) in $\mathbb{R}^3$ and Fig. 19(b) shows a projection of $\mathcal{P}_3$ in $\mathbb{R}^2$, in the plane $\left(r_z, t_{M-y}\right)$.

### 6.3.2 Determination of $\mathcal{P}_2 \cap \mathcal{P}_3$

We shall apply the method described in §5 to determine $\mathcal{P}_2 \cap \mathcal{P}_3$.

Let us define a circumscribed right parallelepiped at the union of the vertices of $\mathcal{P}_2$ and $\mathcal{P}_3$: see Fig. 20.

The intersection of this parallelepiped with the non-cap half-spaces of $\mathcal{P}_2$ and $\mathcal{P}_3$ generates $\mathcal{P}_2 \cap \mathcal{P}_3$: see Fig. 20.

This example shows that if $\mathcal{P}_2 \cap \mathcal{P}_3$ is calculated with a conventional intersection algorithm, the result will be different and will not correspond to the expected result in a tolerance analysis problem.

### 6.3.3 Determination of $\mathcal{P}_1 \cap \left(\mathcal{P}_2 \cap \mathcal{P}_3\right)$

The determination of $\mathcal{P}_1 \cap \left(\mathcal{P}_2 \cap \mathcal{P}_3\right)$ is an unusual case, where the intersection of the non-cap half-spaces of the operands generates a bounded set. The resulting polytope contains no cap half-spaces, see Fig. 21.

In this case, the intersection of a circumscribed right hyper-parallelepiped at the union of the vertices of operands $\mathcal{P}_1$ and $\left(\mathcal{P}_2 \cap \mathcal{P}_3\right)$ with the non-cap half-spaces of these operands gives the same result as the intersection of the non-cap half-spaces: see Fig. 21.

### 6.3.4 Determination of $\mathcal{P}_{3,4/1,4}$

Determination of $\mathcal{P}_{3,4/1,4}$ is shown in Fig. 22. Note that polytopes $\mathcal{P}_{3,4/2,4}$ and $\mathcal{P}_{2,4/2,0}$ are identical as $t_{2,4} = J_4$ according to the hypotheses put forward at the beginning of §6.3. All the operand polytopes $\left(\mathcal{P}_{3,4/2,4}, \mathcal{P}_{2,4/2,0}, \mathcal{P}_{1,0/1,1}, \mathcal{P}_1, \mathcal{P}_2 \text{ and } \mathcal{P}_3\right)$ and hence polytope $\mathcal{P}_{3,4/1,4}$ are centered at the origin of the displacements.

By applying the method described in §4 it can be seen that polytope $\mathcal{P}_{3,4/1,4}$ is made up of two cap polyhedra shown in wireframe in Fig. 22. This means that it is not possible to control geometric variations at point M in translation along direction $\mathbf{x}$ between surfaces 3,4 and 1,4. This specific feature is due to:

- the mobility in translation along $\mathbf{x}$ between parts 3 and 1,
- the displacements in translation along $\mathbf{x}$ leaving surfaces 3,4 and 1,4 invariant.

Fig. 23 shows a calculated polytope $\mathcal{P}_{3,4/1,4}$ included in the functional polyhedron defined in equation (33). Using the descriptors 'cap' and 'non-cap' for the vertices and facets of polytopes in operations applied here it is clear that this inclusion does not depend on cap half-spaces. From this we deduce that the position variations at point M in translation along $\mathbf{y}$ between surfaces 3,4 and 1,4 can be controlled. As a result, the functional condition FC defined in equation (33) is respected.

## 7 Conclusion – future prospects

In the first part of this article, we showed how geometric constraints characterizing the displacements of a surface within a tolerance zone can be made compliant with a geometric polyhedron. In the same way, contact constraints characterizing relative displacements between two surfaces potentially in contact can be made compliant with a contact polyhedron. Cap half-spaces were introduced to be able to delimit geometric polyhedra to transform them into geometric polytopes. These cap half-spaces limit the unbounded displacements of a surface within a tolerance zone. In the same way, contact polyhedra were delimited by cap half-spaces to give contact polytopes. These half-spaces limit the unbounded displacements between two surfaces potentially in contact.

Thus the relative position of any two surfaces of a mechanical system can be formalized by operations, Minkowski sum and intersection, on operand polytopes of $\mathbb{R}^6$. The reason for using cap half-spaces to delimit polyhedra so they become polytopes is related to the algorithmic complexity of Minkowski sums of polyhedra of $\mathbb{R}^n$.

In the second part the mechanisms to identify the cap half-spaces of a Minkowski sum and an intersection were described. The influence of cap half-spaces on the topology of a polytope obtained by summation or intersection is paramount to validate the compliance of a mechanical system in terms of a functional requirement.

The third part demonstrated how operations on polytopes are carried out to simulate compliance of a simple mechanism in terms of a functional condition.

Although the sample application is treated as a plane problem where only polytopes of $\mathbb{R}^3$ could be handled, the mechanisms to identify cap half-spaces in the sums and intersections can be used in $\mathbb{R}^6$ with no restrictions.

A prototype of tolerance analysis software is currently being developed based on the concepts of algorithmic geometry presented in this article. With this tool it will be possible to calculate the sums and intersections of polytopes of $\mathbb{R}^6$ and ensure that cap half-spaces are identified. In future it should be possible to use this tool to experiment on complex mechanical systems.

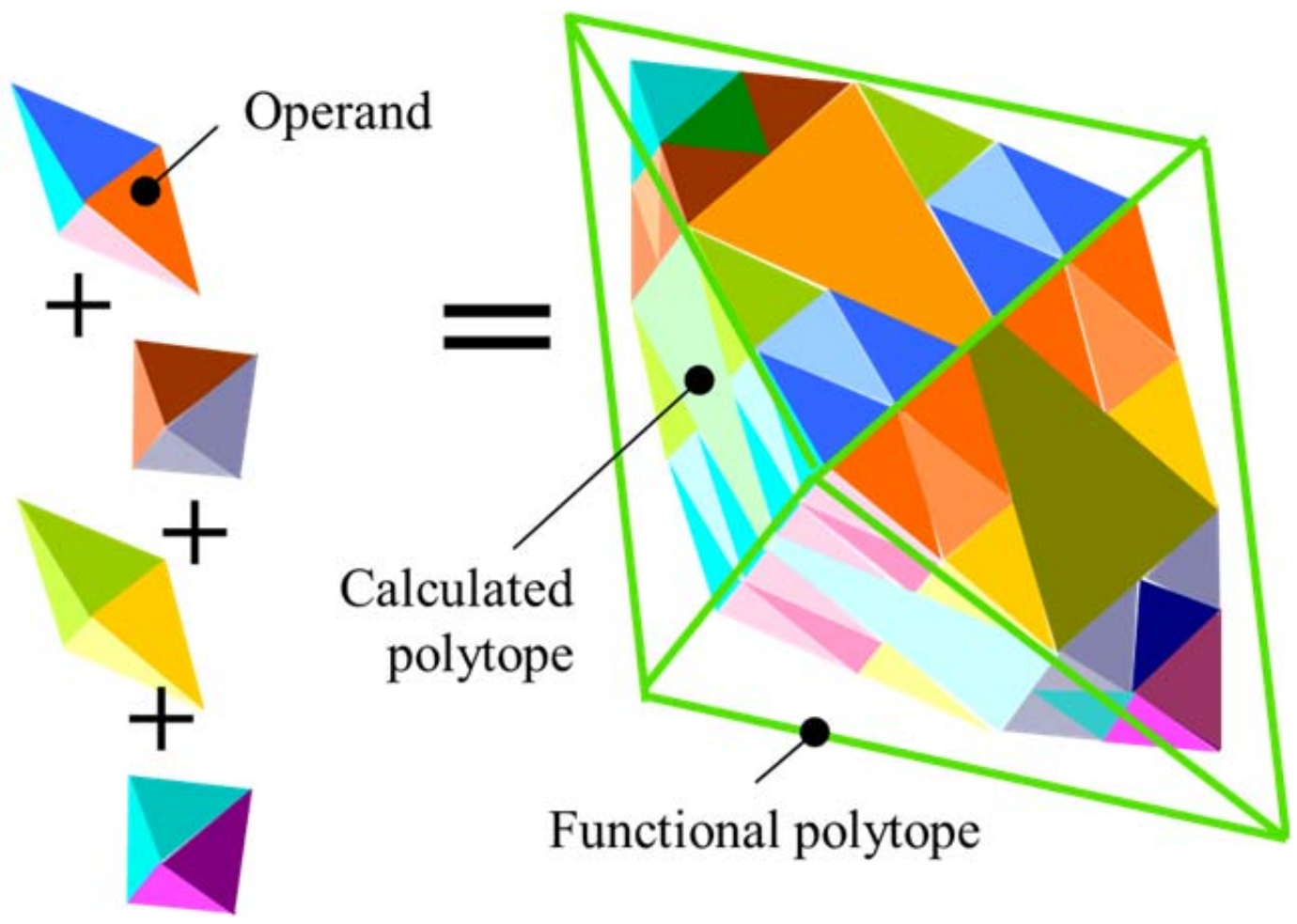

Fig. 1. Verifying compliance with functional requirement [4].

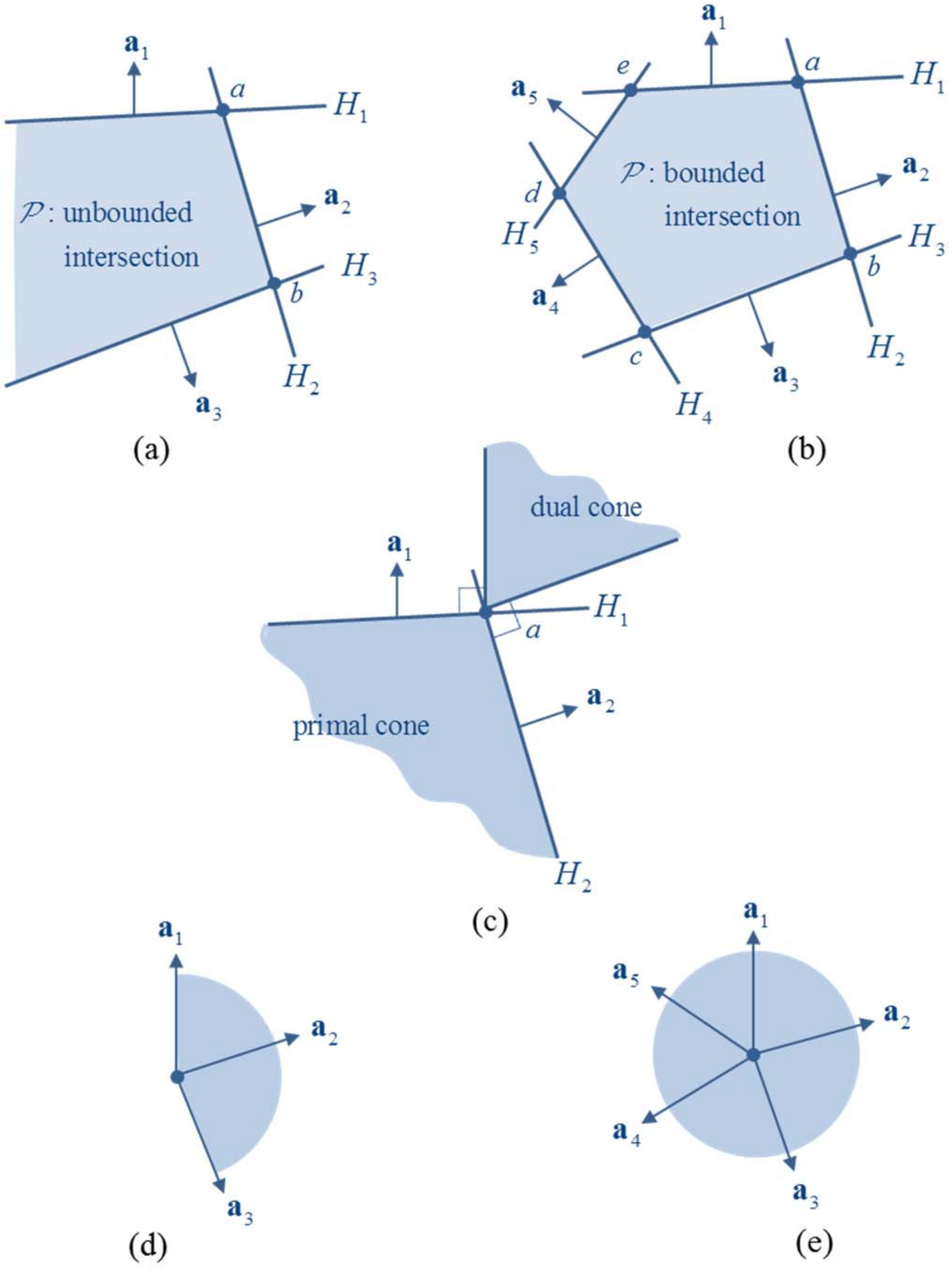

Fig. 2. Polyhedron, polytope and normal fans.

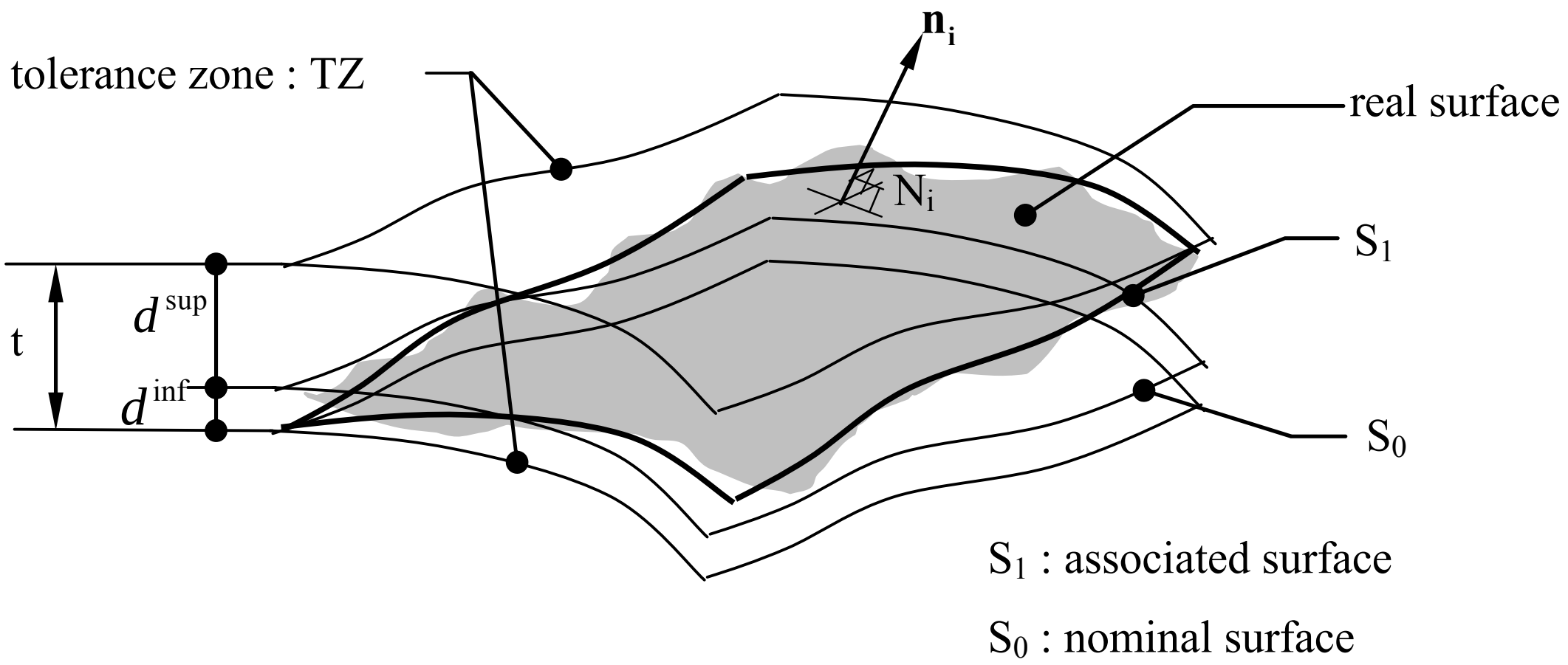

Fig. 3. Definition of a tolerance zone.

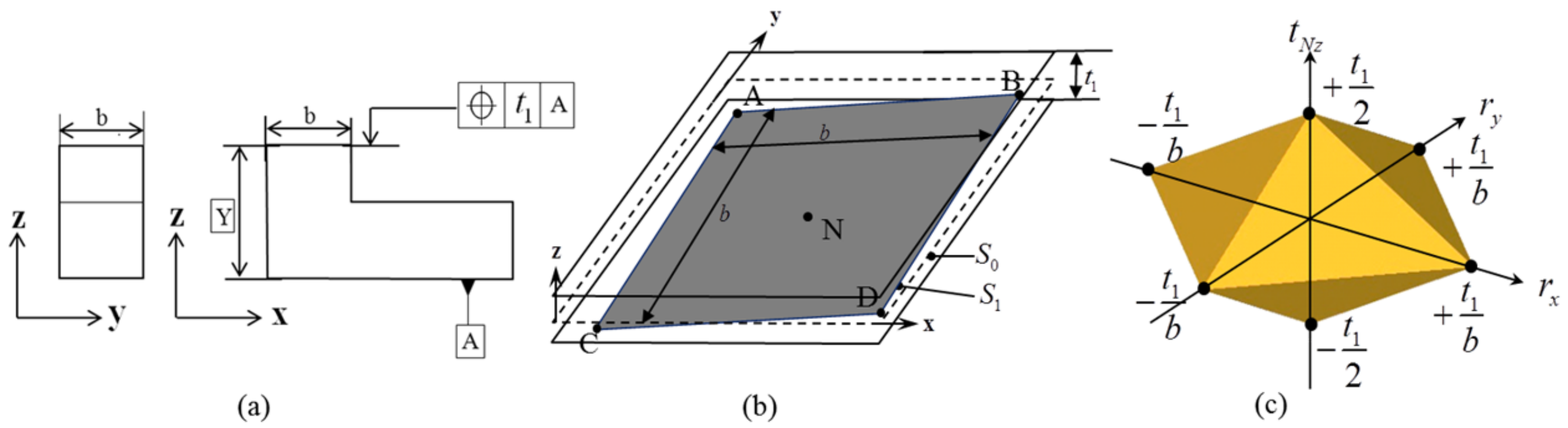

Fig. 4. Geometric constraints of a plane surface within a tolerance zone.

| Surfaces in contact | Contact element |
|---|---|
| Planar pair (Plane/plane)<br>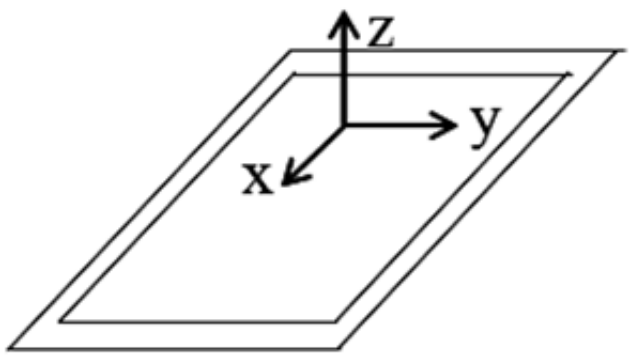 | Plane *P*<br>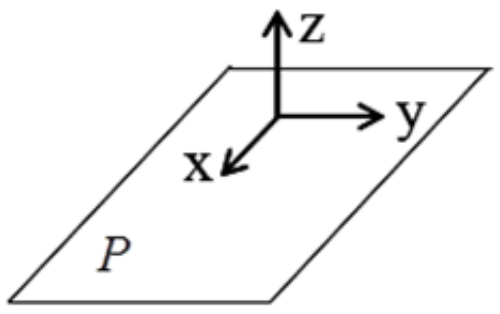 |
| Cylindrical pair (Cylinder/cylinder)<br>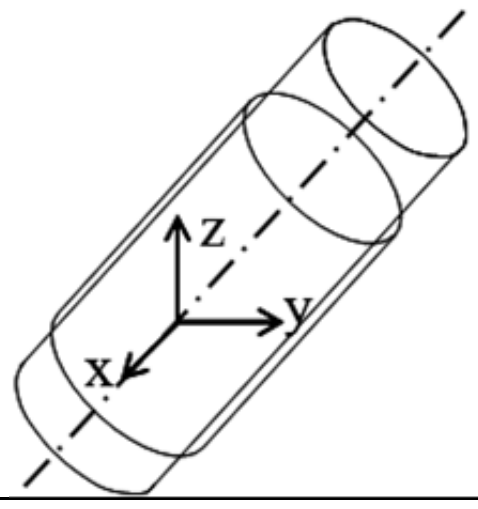 | Cylinder *C*<br>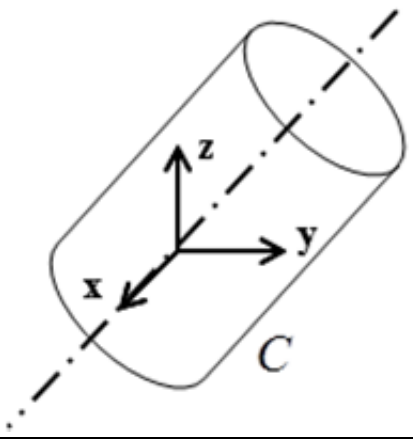 |
| Cylinder-and-plane pair (Plane/cylinder)<br>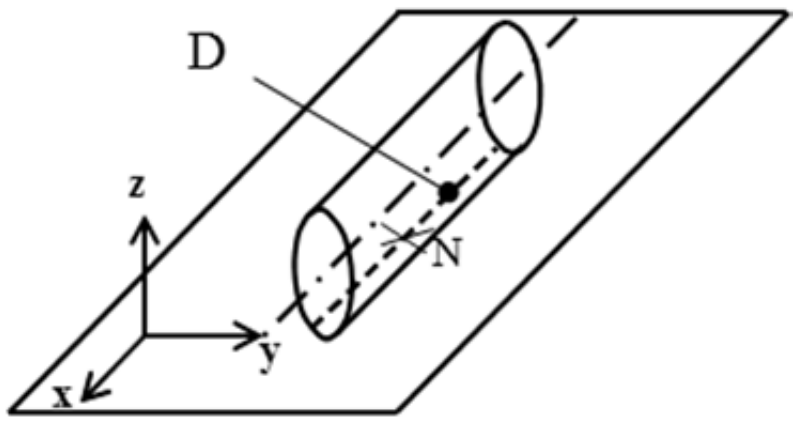 | Segment of straight line *D* |
| Ball-and-plane pair (Plane/sphere)<br>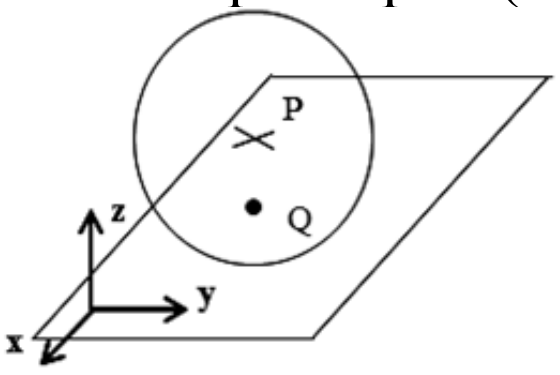 | Point *Q* |

Fig. 5. Examples of contact elements.

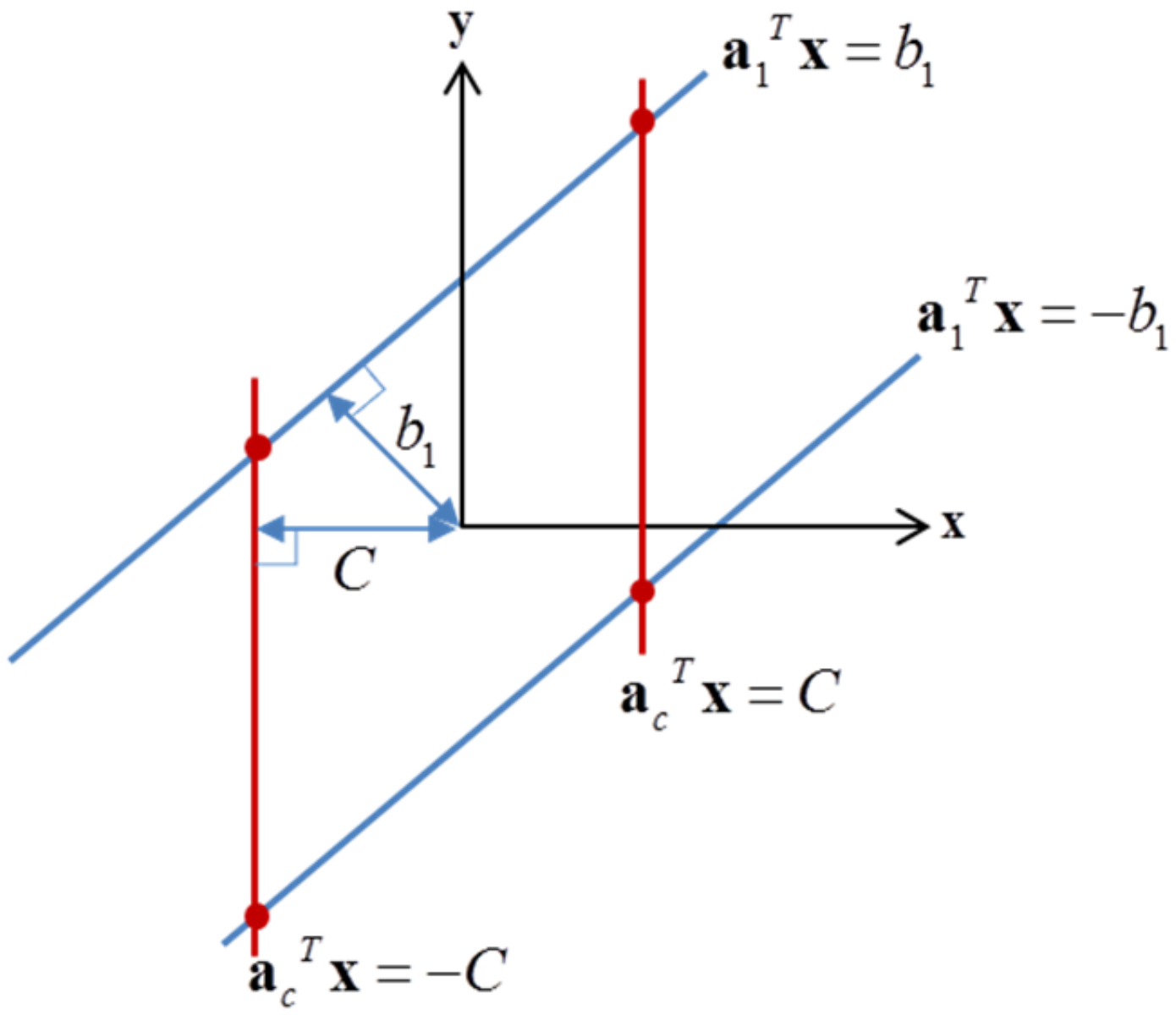

Fig. 6. Definition of cap half-spaces.

| **Class of surface** | **Degree of invariance - displacements leaving surface invariant** | **Cap half-spaces** |
|---|---|---|
| Plane 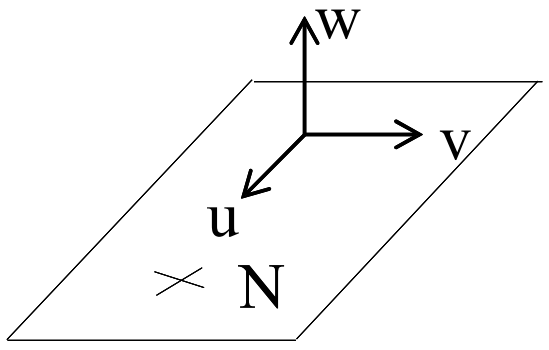  | Degree of invariance = 3 $\begin{cases} \mathbf{r}.\mathbf{w} \\ \mathbf{t}_N.\mathbf{u} \\ \mathbf{t}_N.\mathbf{v} \end{cases}$ | $\begin{cases} \mathbf{r}.\mathbf{w} \le C & -\mathbf{r}.\mathbf{w} \le C \\ \mathbf{t}_N.\mathbf{u} \le C & , -\mathbf{t}_N.\mathbf{u} \le C \\ \mathbf{t}_N.\mathbf{v} \le C & -\mathbf{t}_N.\mathbf{v} \le C \end{cases}$ |
| Cylindrical 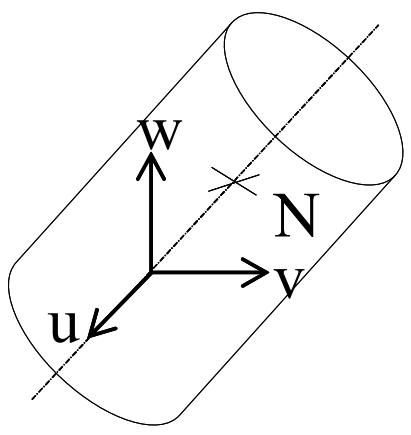  | Degree of invariance = 2 $\begin{cases} \mathbf{r}.\mathbf{u} \\ \mathbf{t}_N.\mathbf{u} \end{cases}$ | $\begin{cases} \mathbf{r}.\mathbf{u} \le C & -\mathbf{r}.\mathbf{u} \le C \\ \mathbf{t}_N.\mathbf{u} \le C & , -\mathbf{t}_N.\mathbf{u} \le C \end{cases}$ |
| Spherical 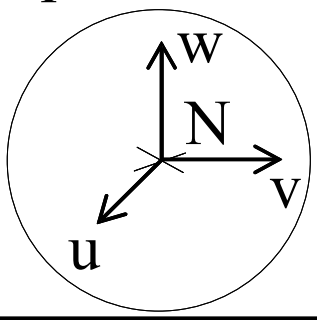  | Degree of invariance = 3 $\begin{cases} \mathbf{r}.\mathbf{u} \\ \mathbf{r}.\mathbf{v} \\ \mathbf{r}.\mathbf{w} \end{cases}$ | $\begin{cases} \mathbf{r}.\mathbf{u} \le C & -\mathbf{r}.\mathbf{u} \le C \\ \mathbf{r}.\mathbf{v} \le C & , -\mathbf{r}.\mathbf{v} \le C \\ \mathbf{r}.\mathbf{w} \le C & -\mathbf{r}.\mathbf{w} \le C \end{cases}$ |
| Of revolution 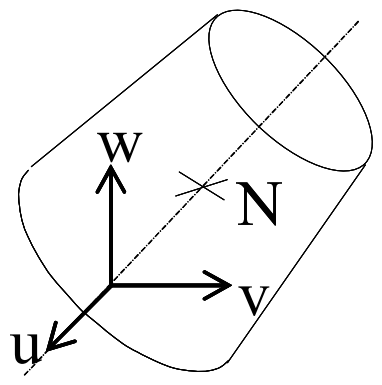  | Degree of invariance = 1 $\mathbf{r}.\mathbf{u}$ | $\mathbf{r}.\mathbf{u} \le C, -\mathbf{r}.\mathbf{u} \le C$ |
| Prismatic 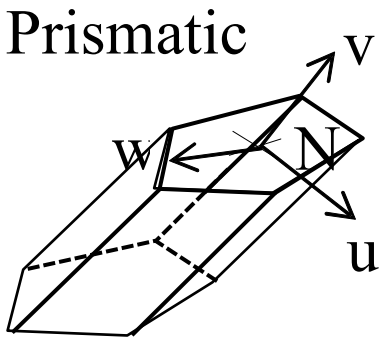  | Degree of invariance = 1 $\mathbf{t}_N.\mathbf{v}$ | $\mathbf{t}_N.\mathbf{v} \le C, -\mathbf{t}_N.\mathbf{v} \le C$ |

Fig. 7. Expression of cap half-spaces for the main classes of surface.

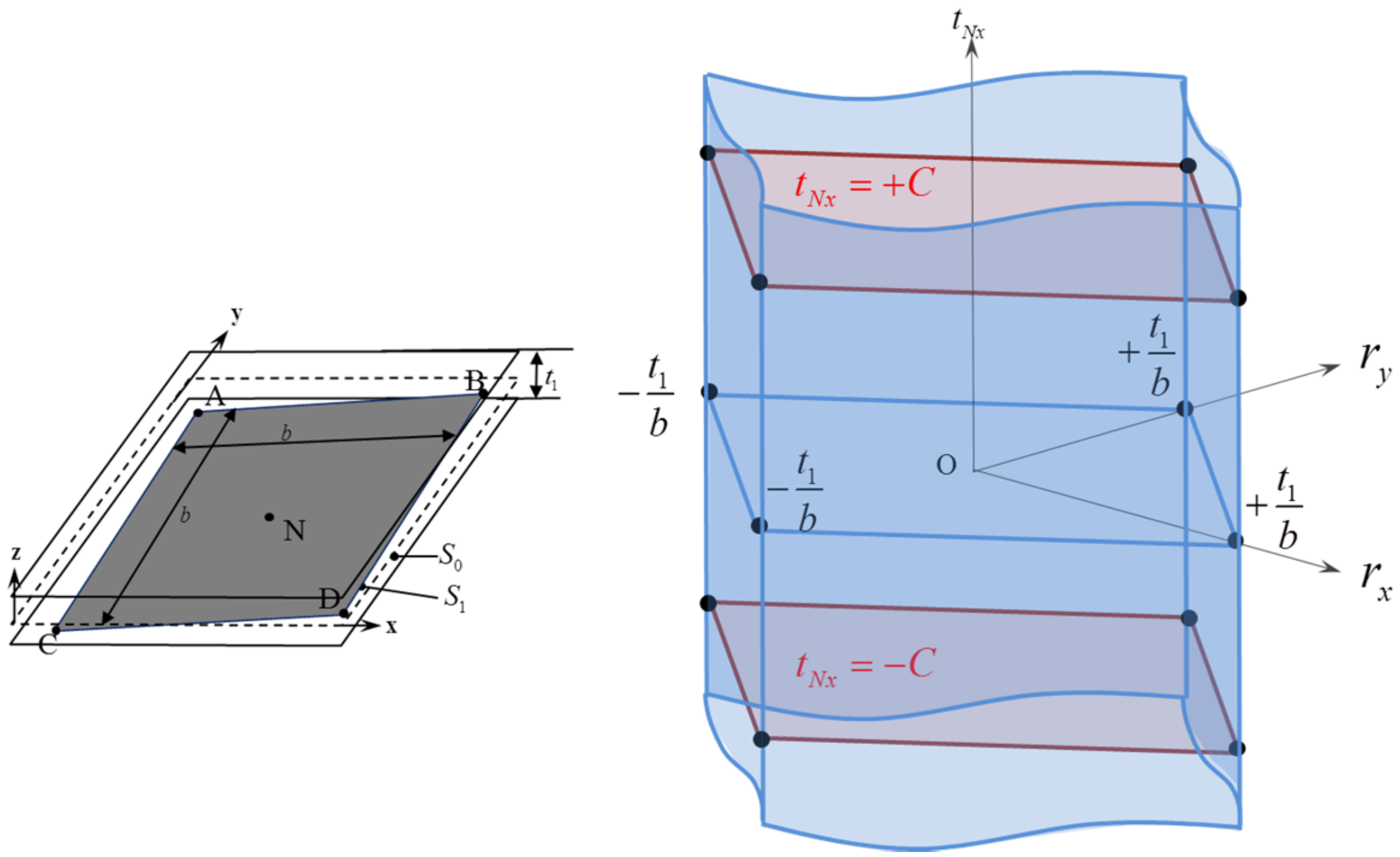

Fig. 8. Unbounded displacement according to $t_{Nx}$.

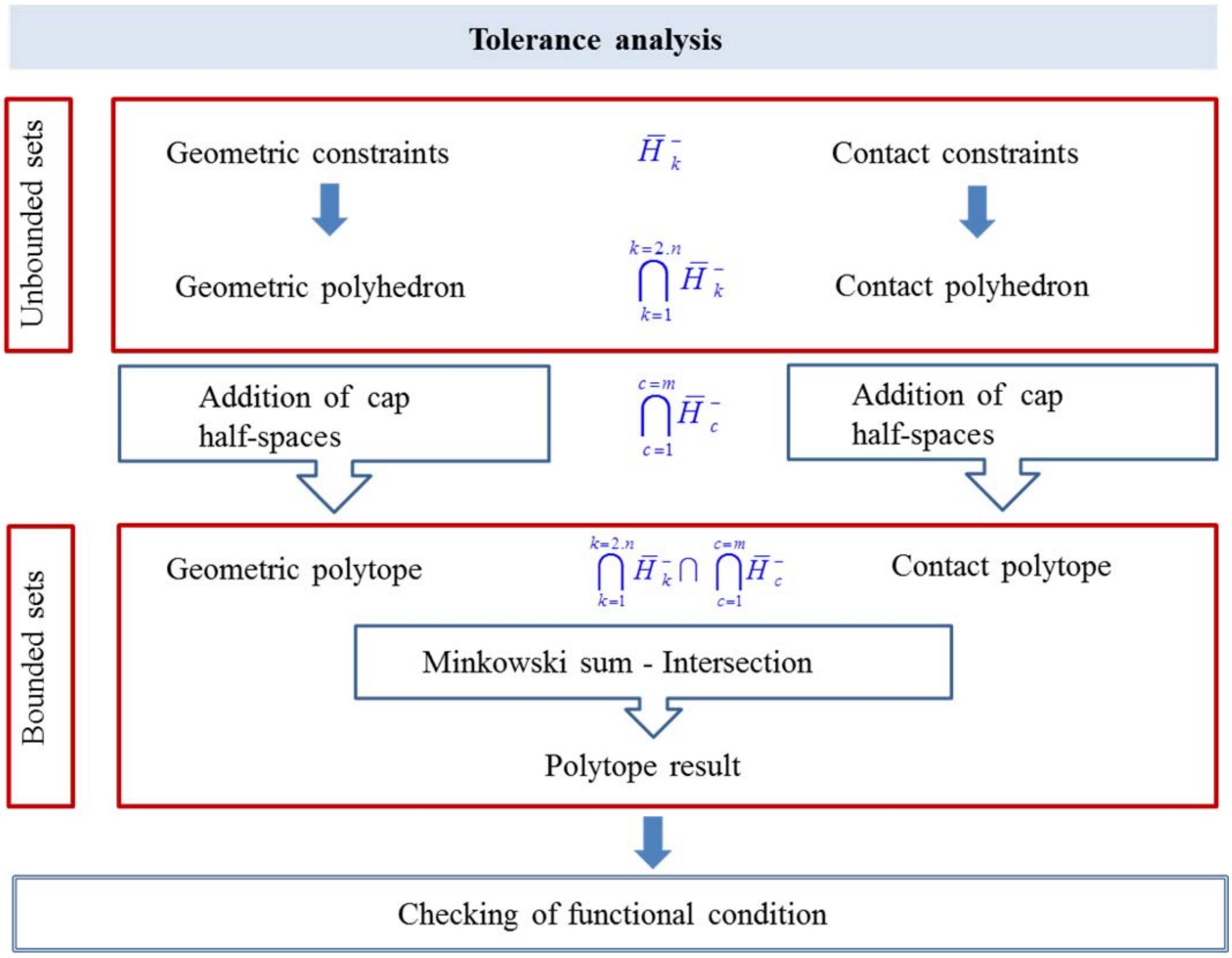

Fig. 9. Synthesis: tolerance analysis using the polytope approach.

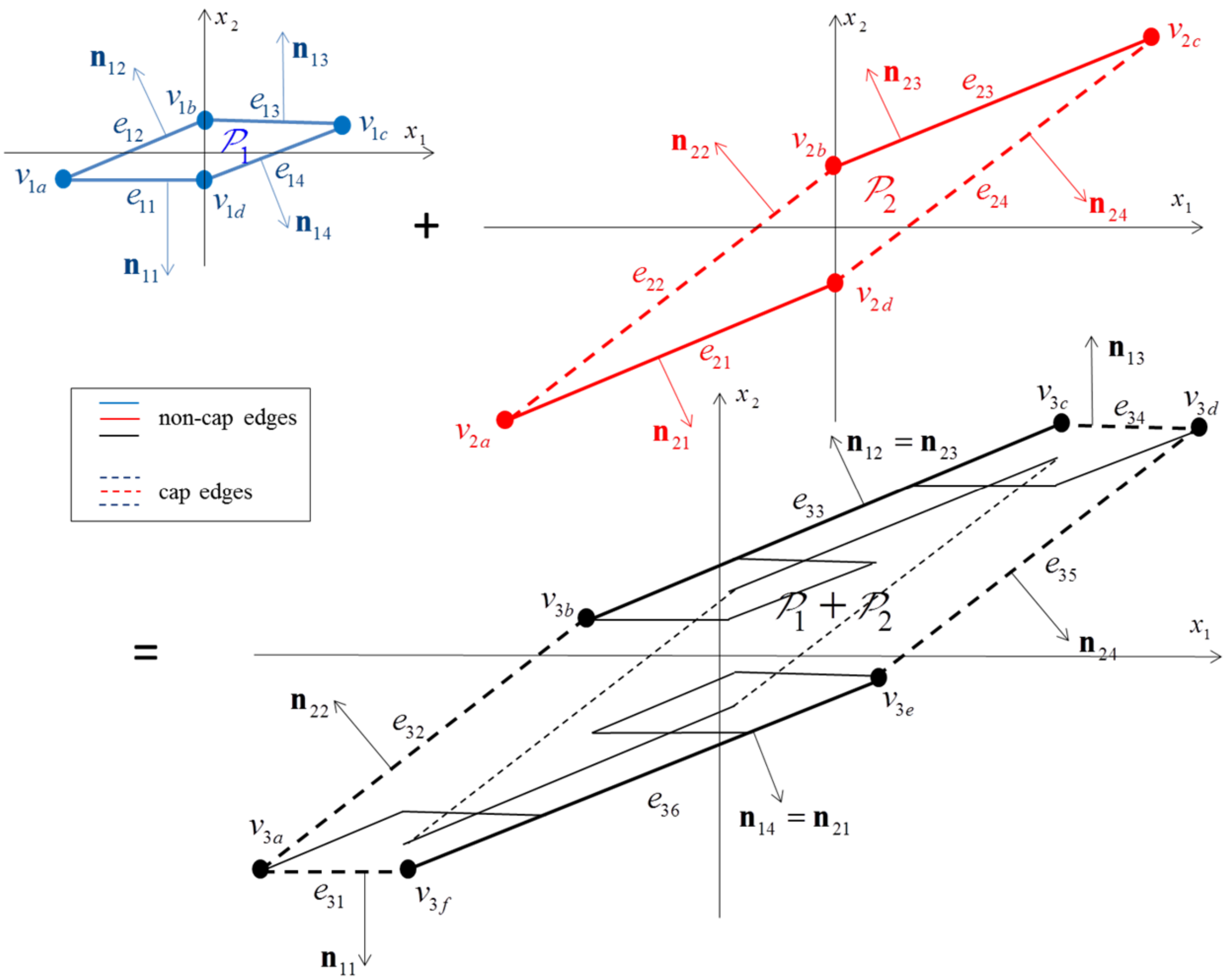

Fig. 10. Minkowski sum for two polytopes in $\mathbb{R}^2$.

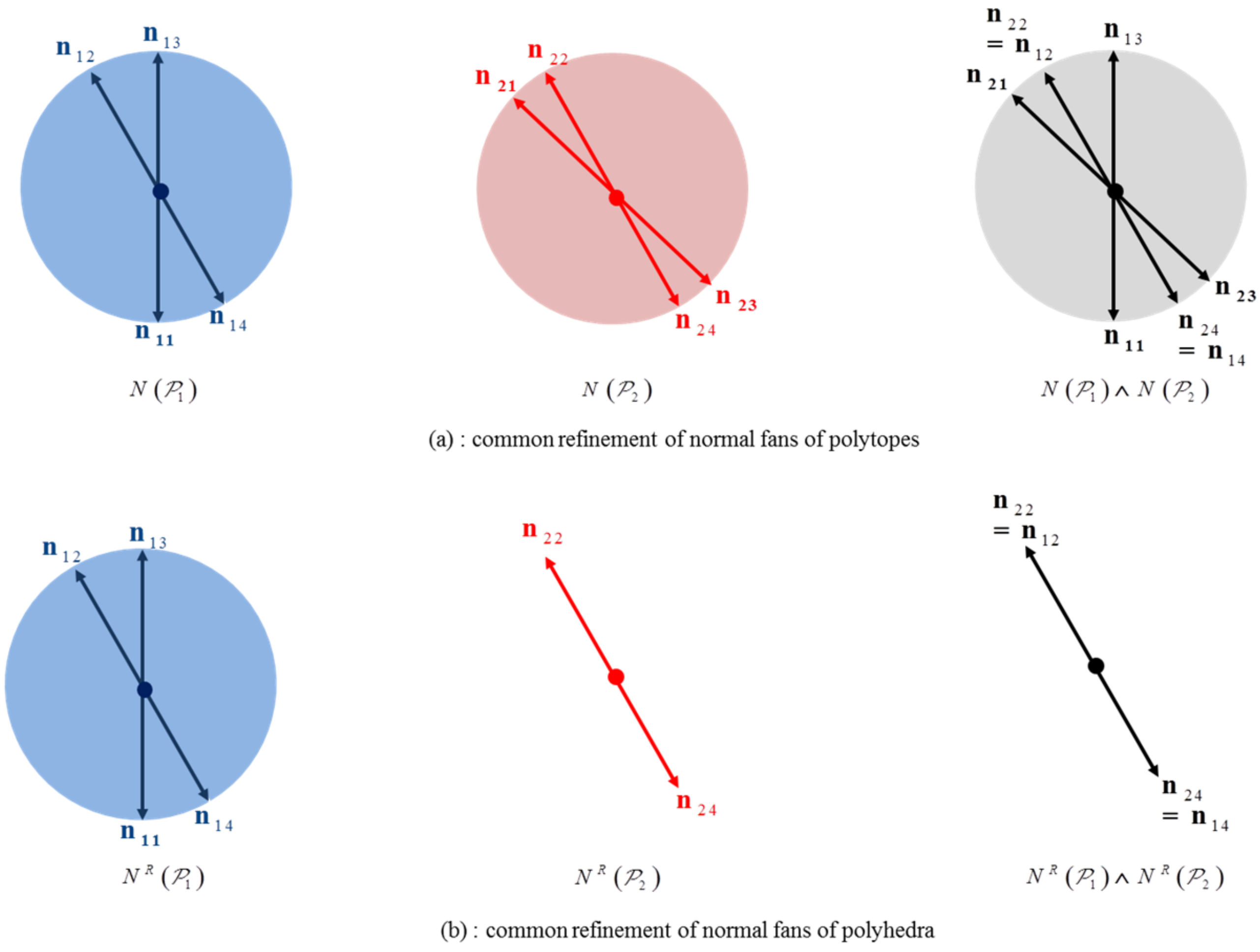

Fig. 11. Common refinement of normal fans of operands.

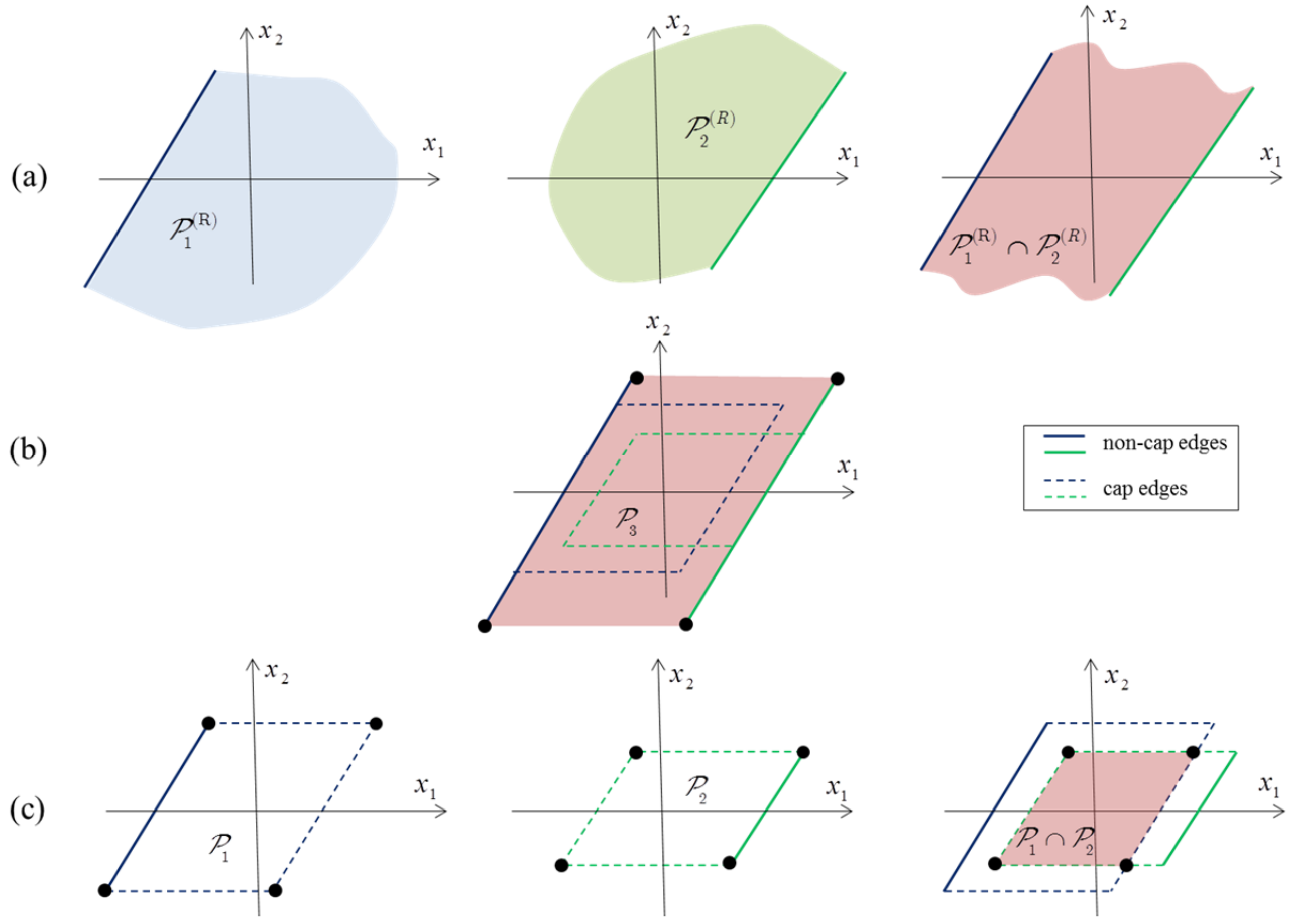

Fig. 12. Intersections of polyhedra $\mathcal{P}_1^{(R)}$ and $\mathcal{P}_2^{(R)}$ and polytopes $\mathcal{P}_1$ and $\mathcal{P}_2$.

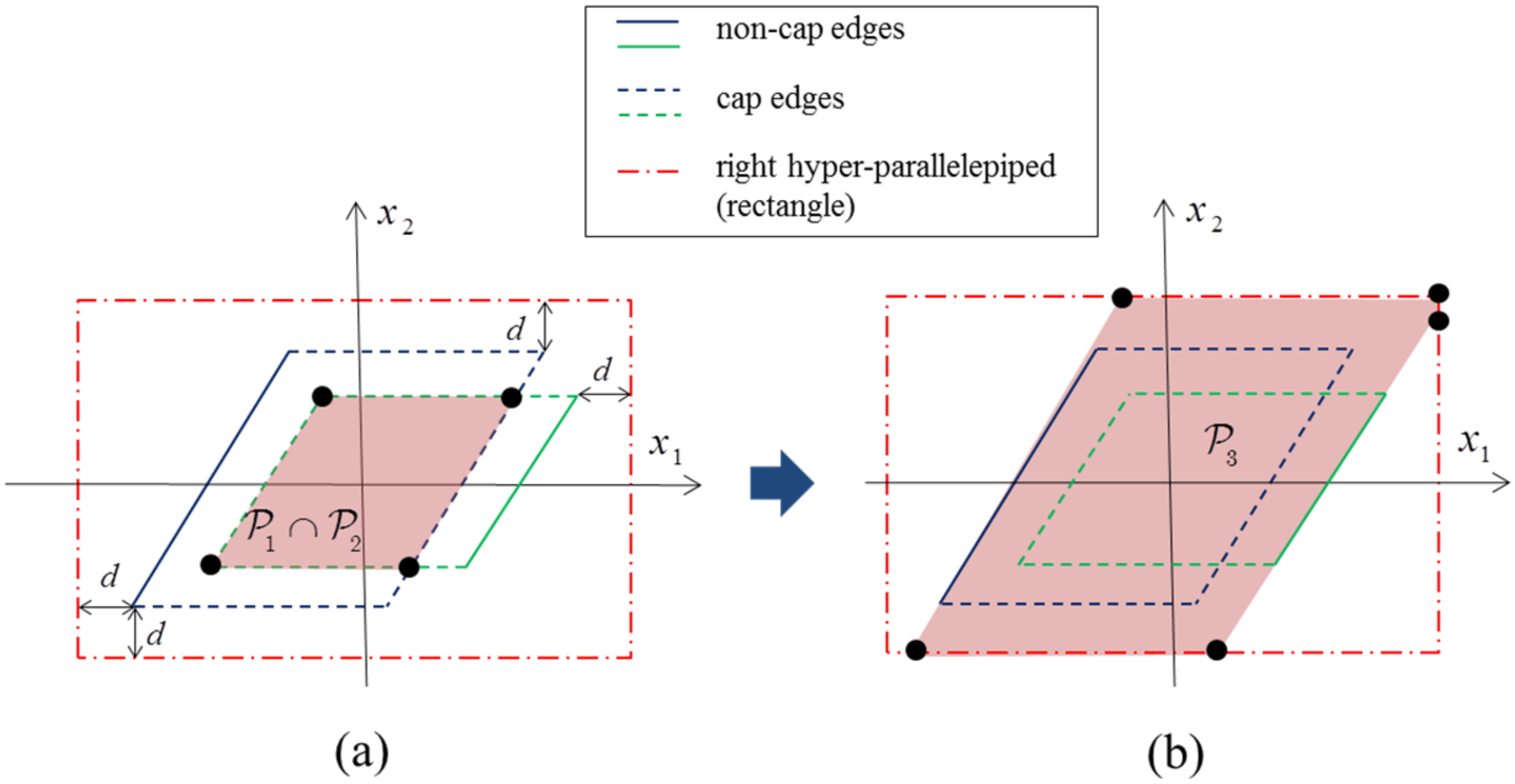

Fig. 13. Determination of $\mathcal{P}_3$.

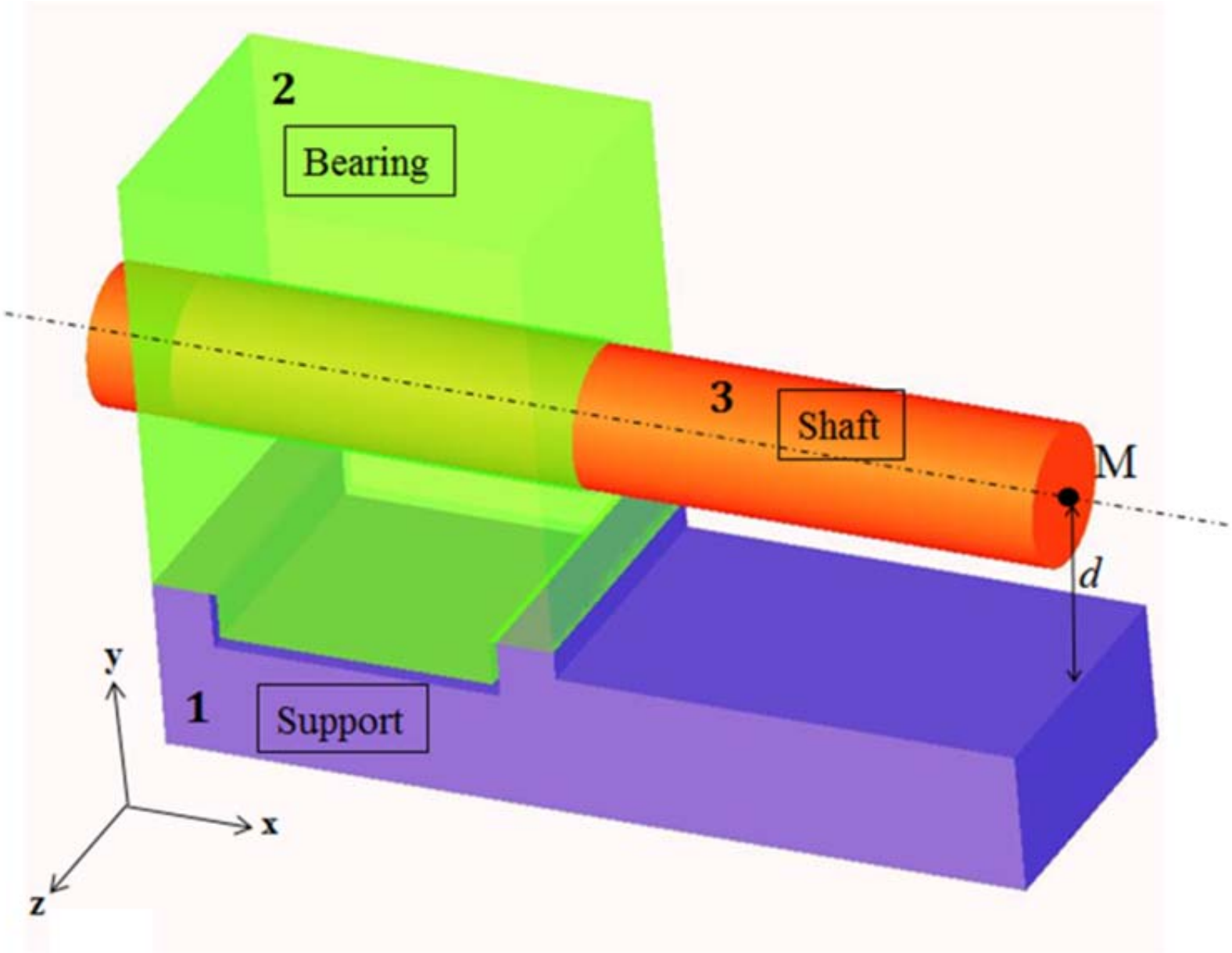

Fig. 14. Mechanism studied.

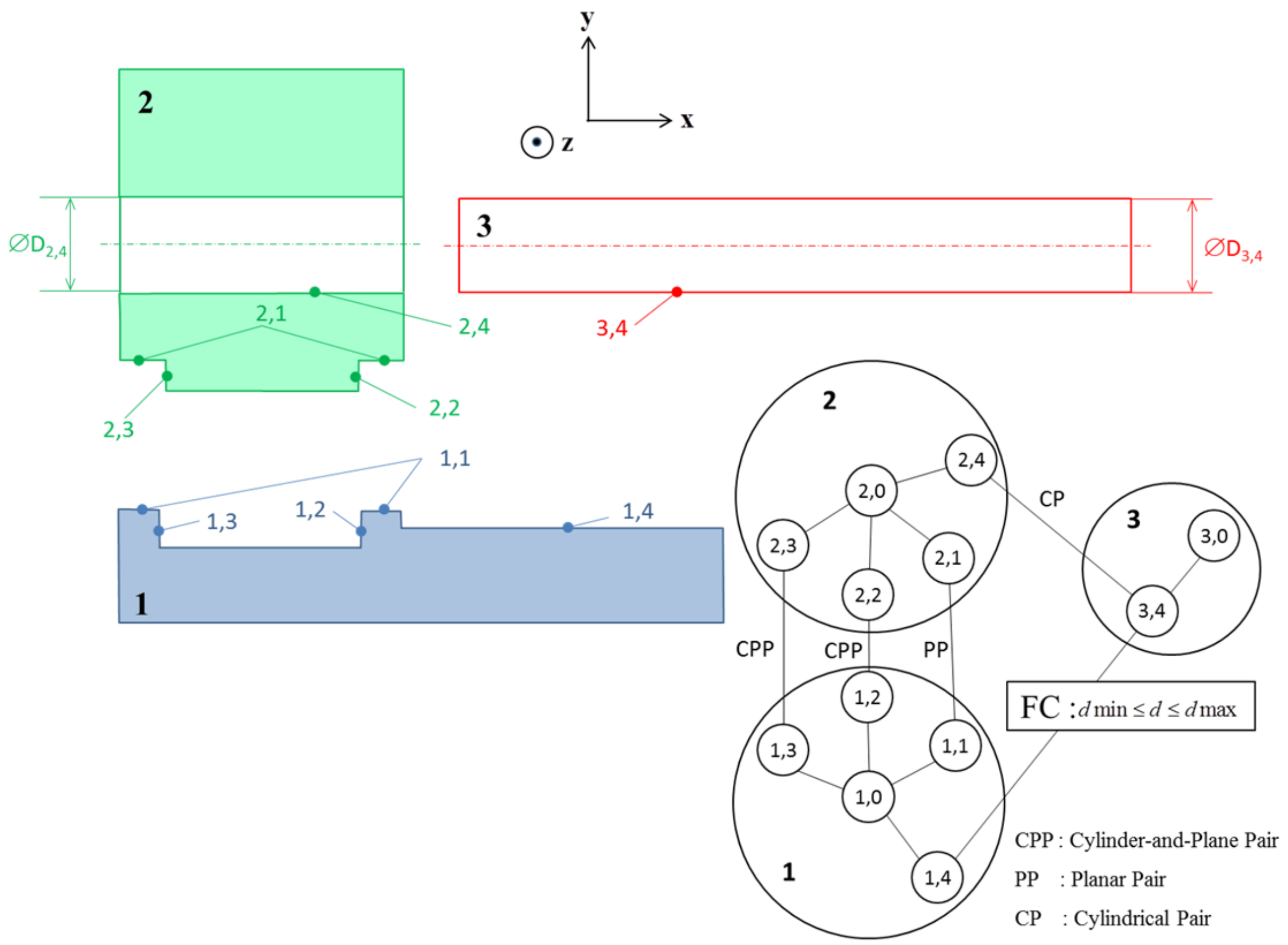

Fig. 15. Definition of the topological structure of the mechanism.

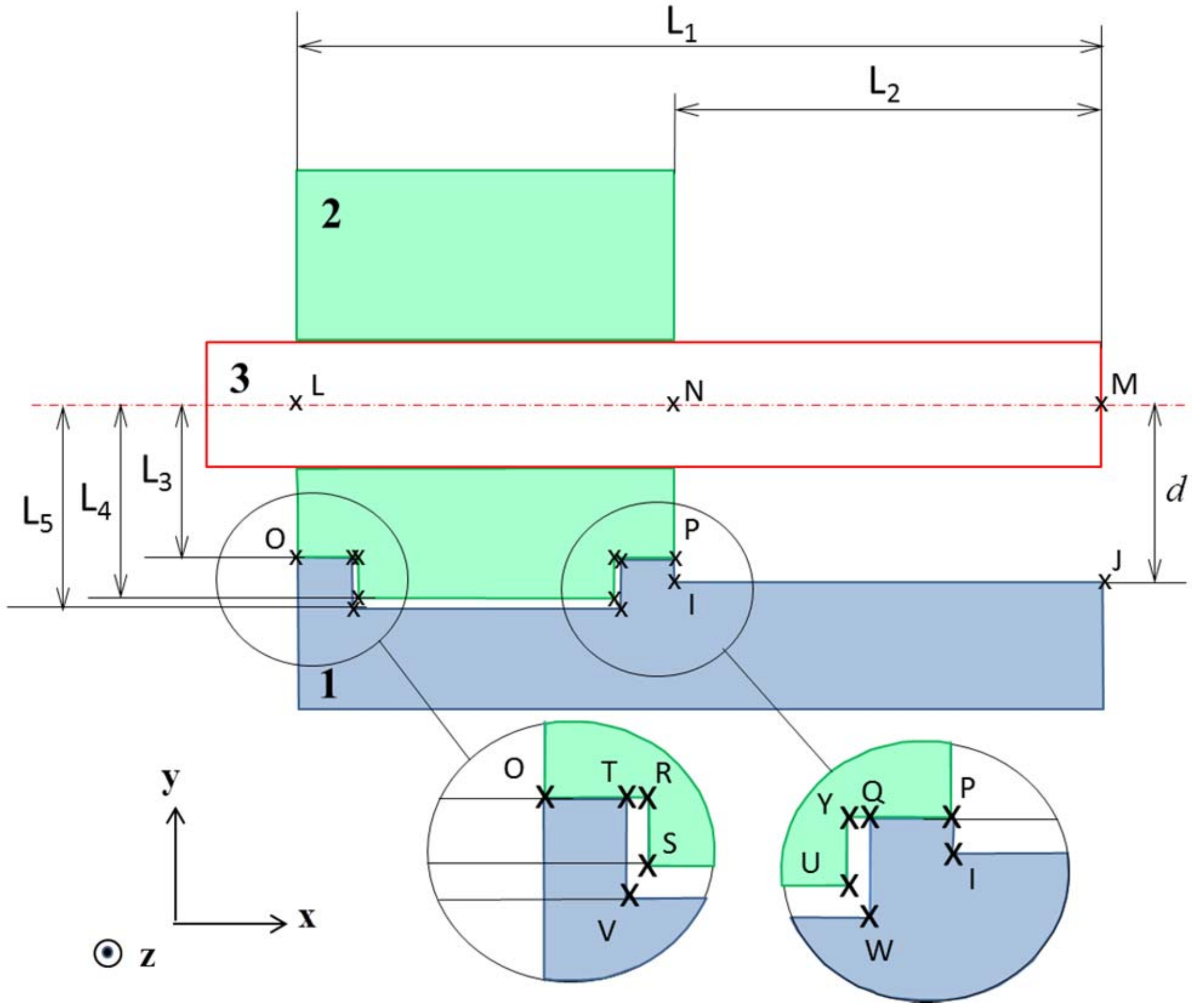

Fig. 16. Definitions of points of expression for geometric constraints, contact constraints and cap half-spaces.

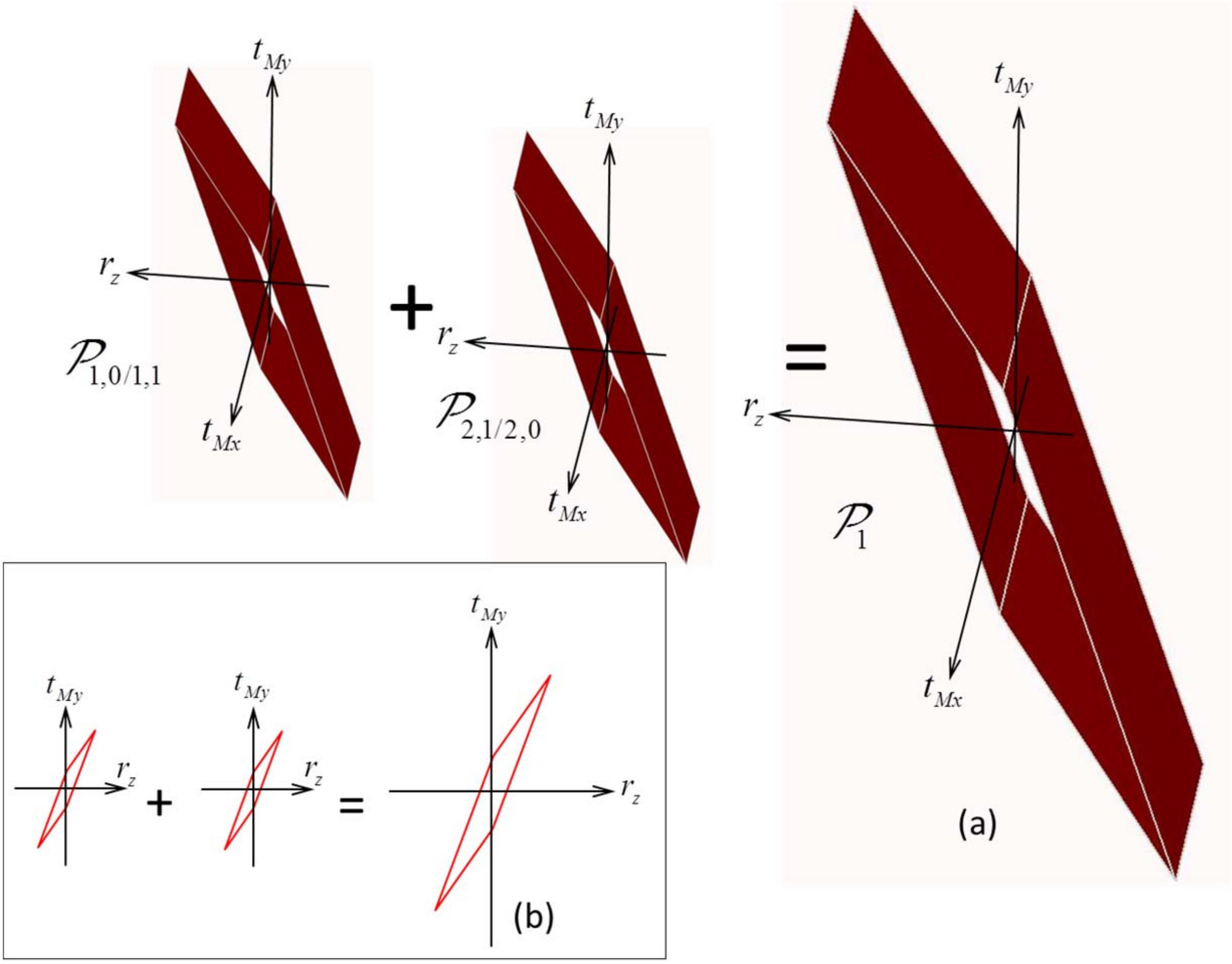

Fig. 17. Determination of polytope $\mathcal{P}_1$.

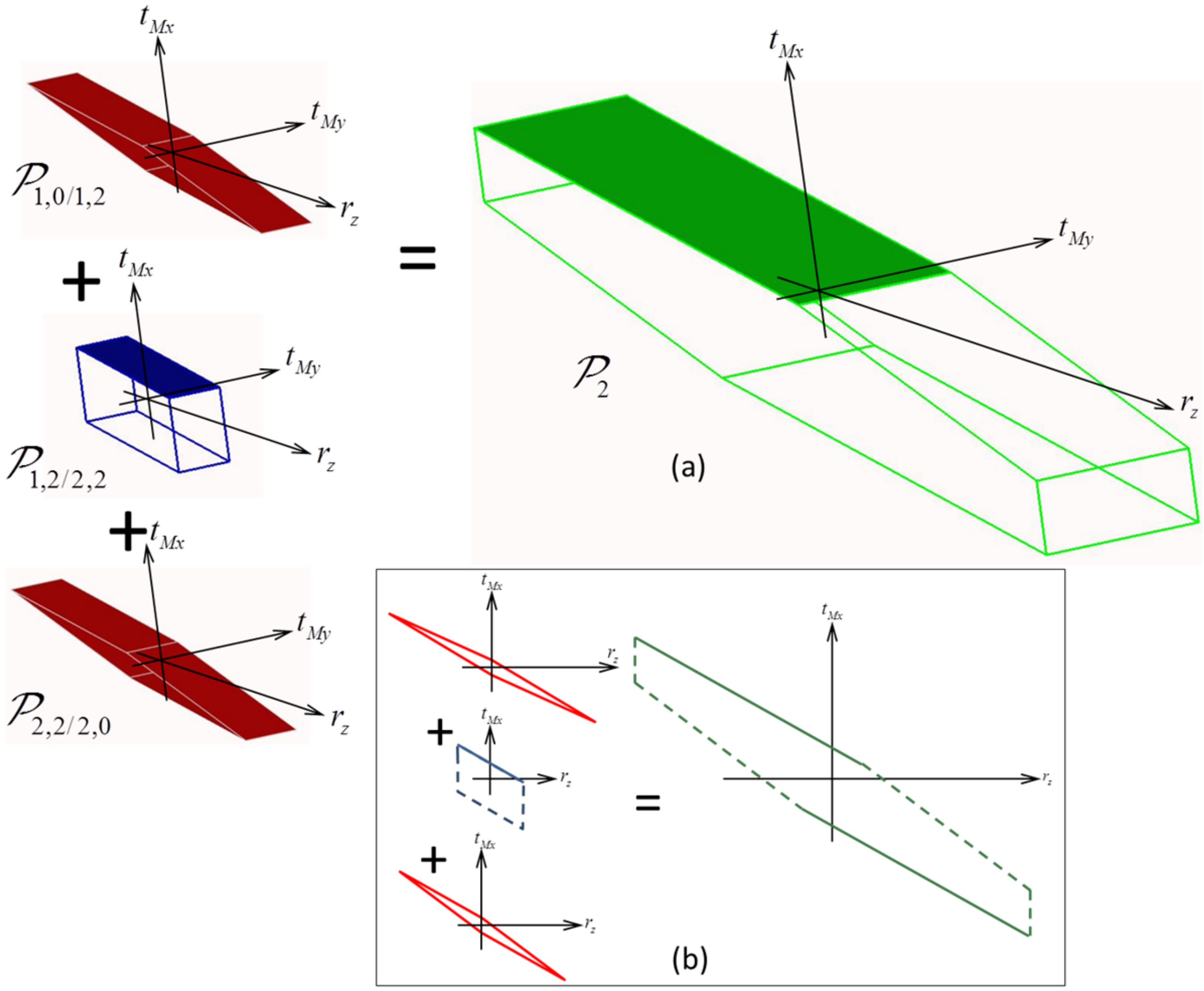

Fig. 18. Determination of polytope $\mathcal{P}_2$.

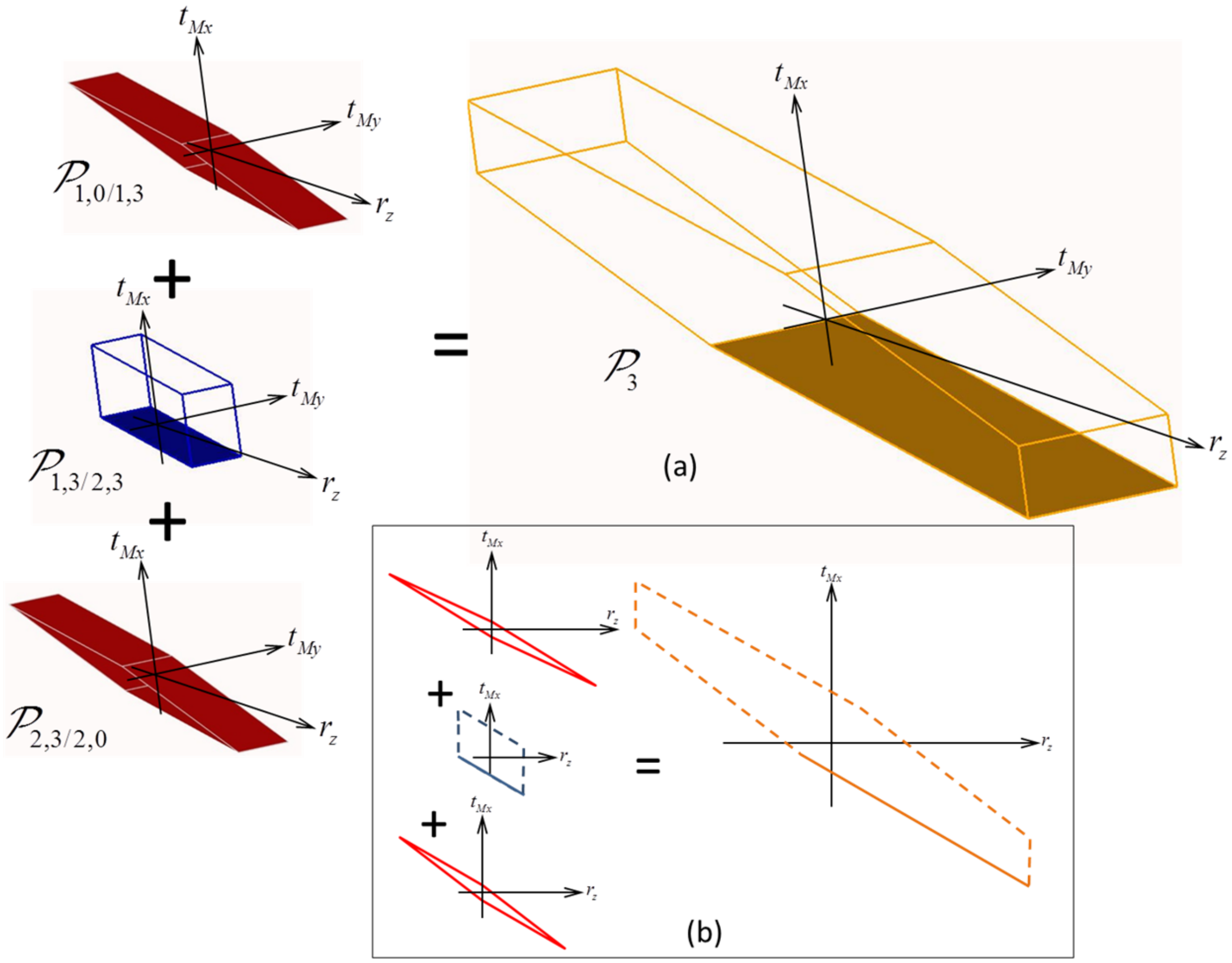

Fig. 19. Determination of polytope $\mathcal{P}_3$.

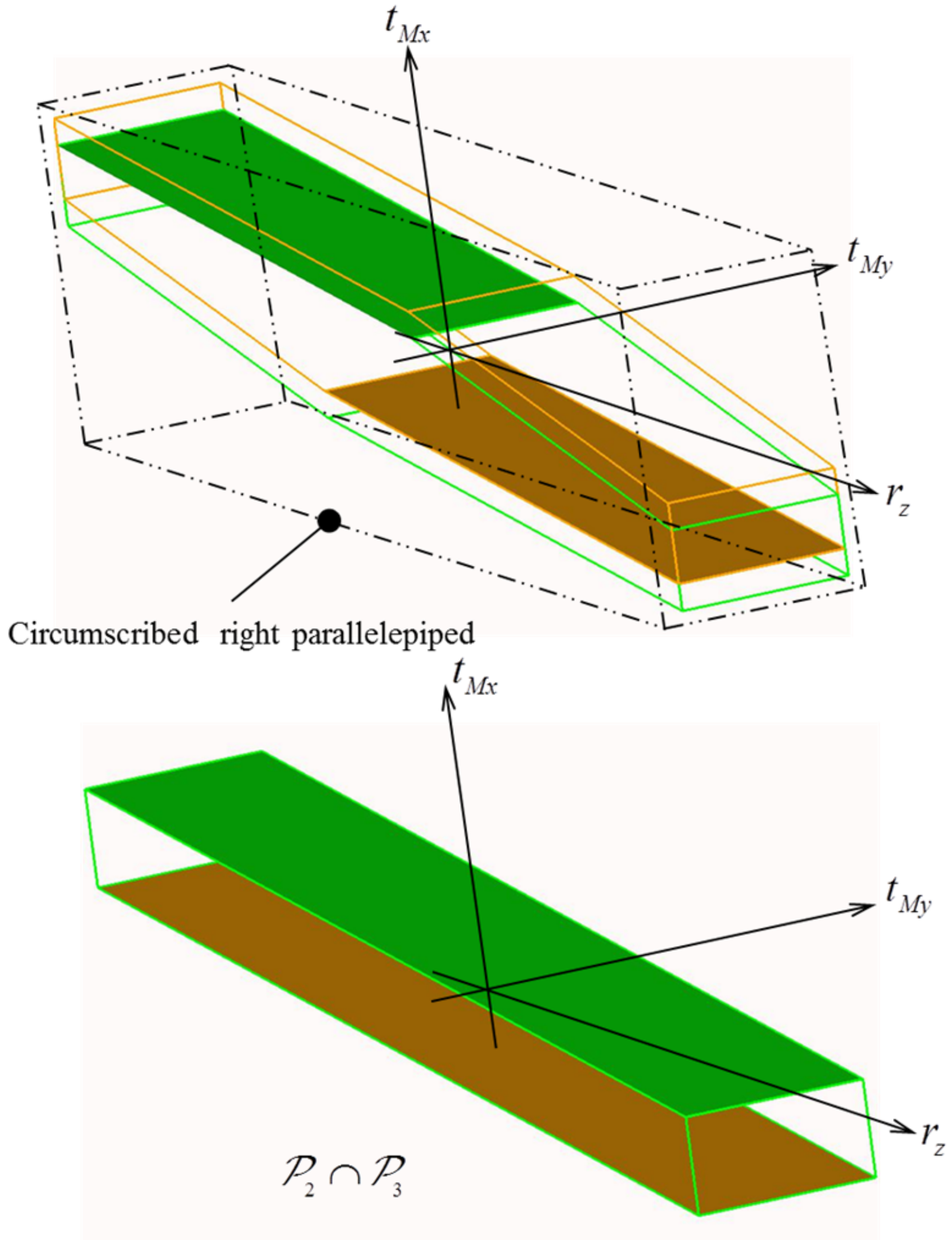

Fig. 20. Determination of $\mathcal{P}_2 \cap \mathcal{P}_3$

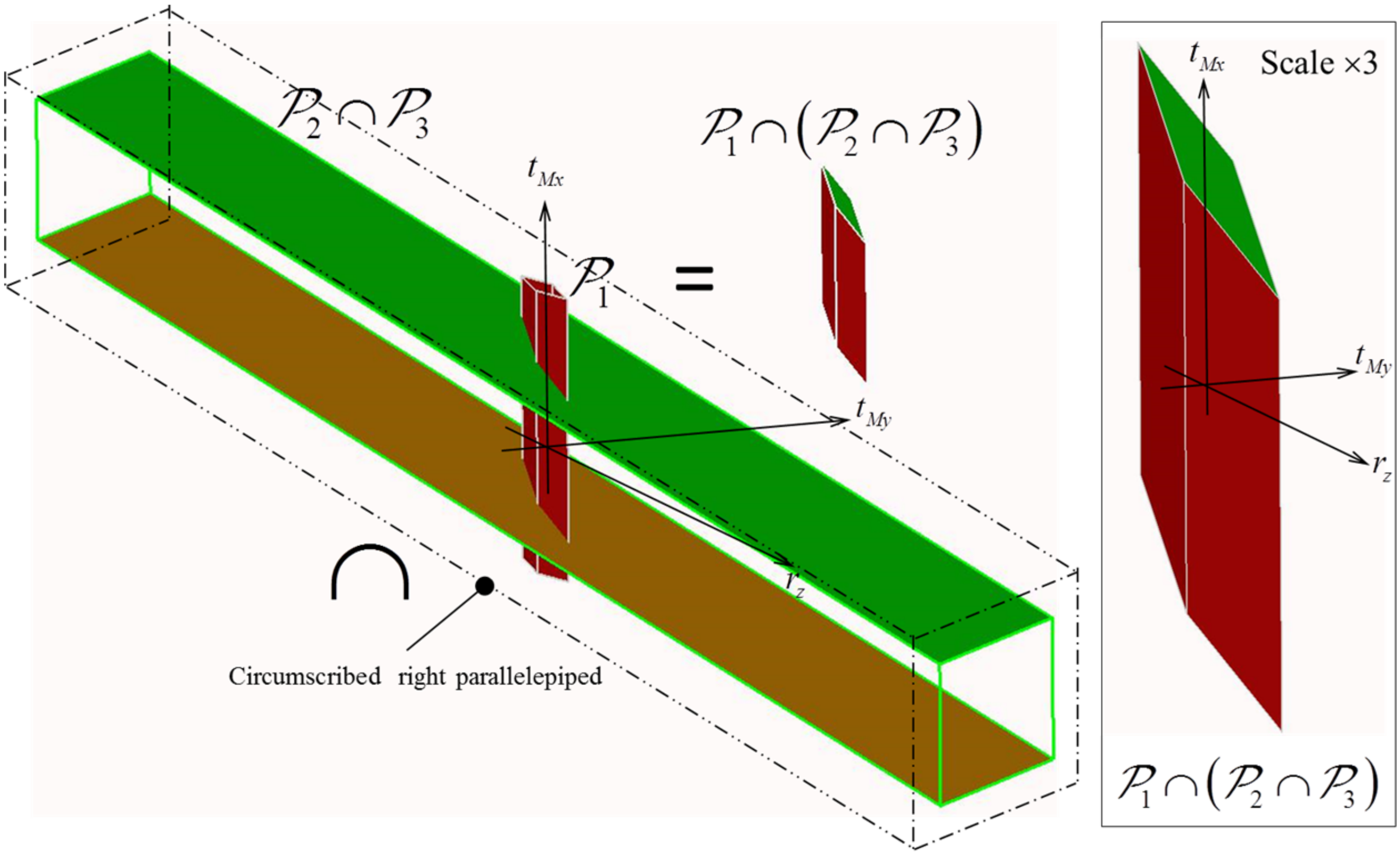

Fig. 21. Determination of $\mathcal{P}_1 \cap (\mathcal{P}_2 \cap \mathcal{P}_3)$

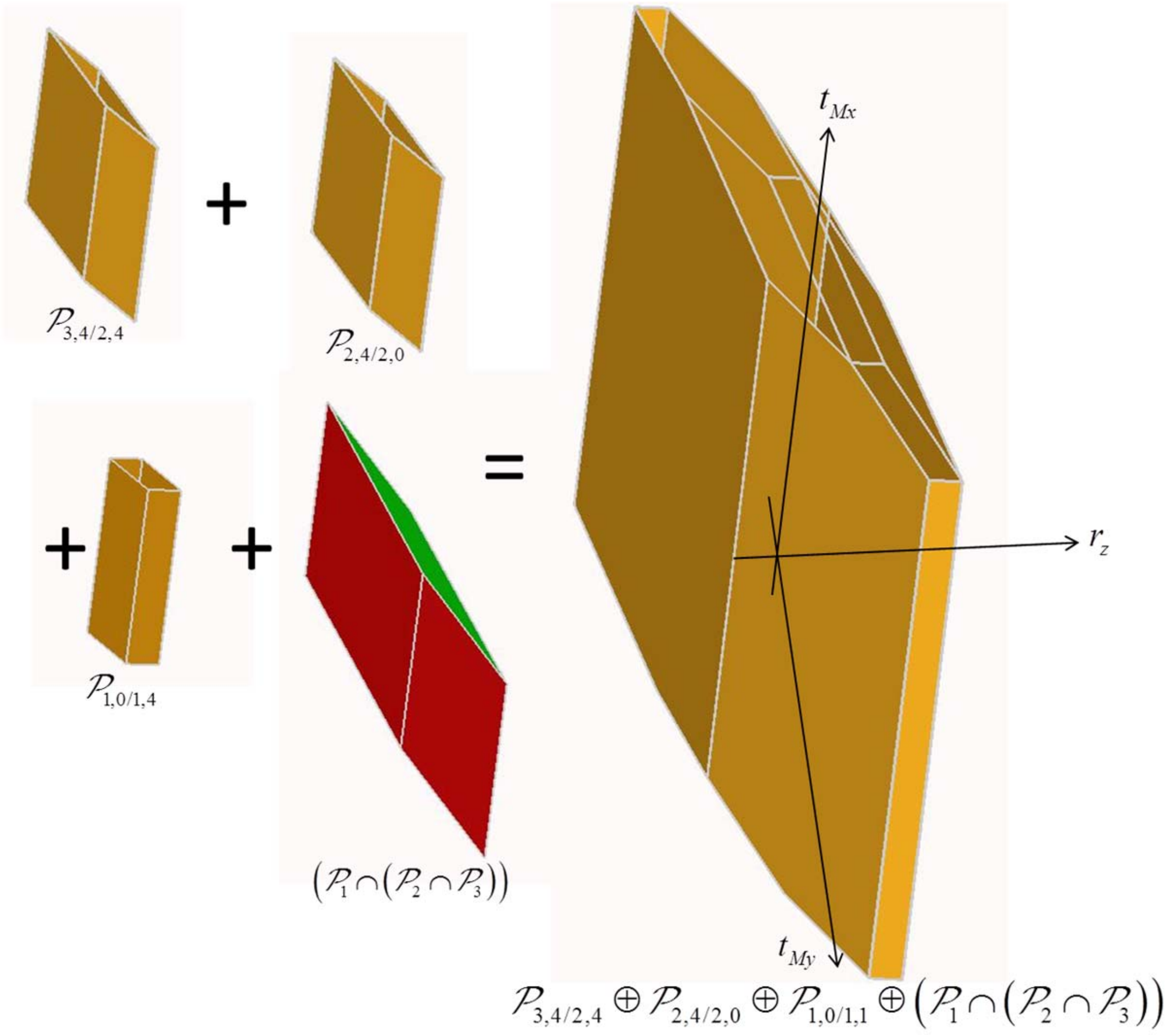

Fig. 22 Determination of $\mathcal{P}_{3,4/1,4} = \mathcal{P}_{3,4/2,4} \oplus \mathcal{P}_{2,4/2,0} \oplus \mathcal{P}_{1,0/1,1} \oplus \left(\mathcal{P}_1 \cap \left(\mathcal{P}_2 \cap \mathcal{P}_3\right)\right)$

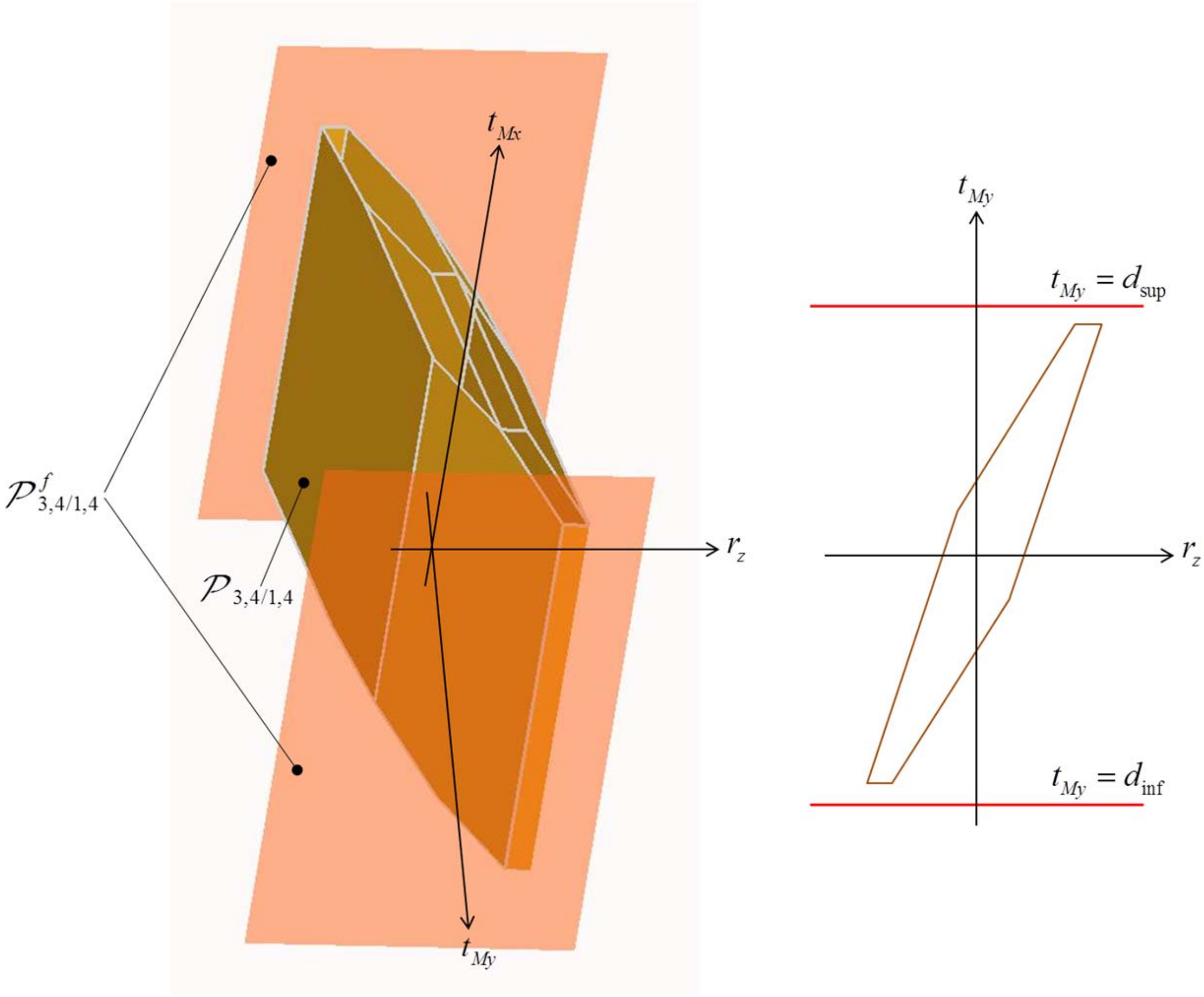

Fig. 23. Inclusion of $\mathcal{P}_{3,4/1,4}$ in the functional polyhedron $\mathcal{P}^f_{3,4/1,4}$.

| **Class of surface** | **Degrees of invariance** | **Number of cap half-spaces** |
|---|---|---|
| **Planar** | 3 | 6 |
| **Cylindrical** | 2 | 4 |
| **Spherical** | 3 | 6 |
| **Prismatic** | 1 | 2 |
| **Revolution** | 1 | 2 |
| **Helicoidal** | 2 | 4 |
| **Complex** | 0 | 0 |

Table 1. Number of cap half-spaces according to class of surface

| Contact | Degrees of freedom | Number of cap half-spaces |
|---|---|---|
| **Planar pair** | 3 | 6 |
| **Cylindrical pair** | 2 | 4 |
| **Ball-and-plane pair** | 5 | 10 |
| **Spherical pair** | 3 | 6 |
| **Ball-and-cylinder pair** | 4 | 8 |
| **Cylinder-and-plane pair** | 4 | 8 |
| **Prismatic pair** | 1 | 2 |
| **Turning pair** | 1 | 2 |

Table 2. Number of cap half-spaces according to the degrees of freedom of the joints.

| | Geometric constraints | Cap half-spaces | Geometric polytope in M |
|---|---|---|---|
| 1,1 / 1,0 | $-\frac{t_{1,1}}{2} \le \mathbf{t}_{\mathrm{O}-1,1/1,0}.\mathbf{y} \le +\frac{t_{1,1}}{2} \perp$<br>$-\frac{t_{1,1}}{2} \le \mathbf{t}_{\mathrm{P}-1,1/1,0}.\mathbf{y} \le +\frac{t_{1,1}}{2}$ | $-C \le \mathbf{t}_{\mathrm{M}-1,1/1,0}.\mathbf{x} \le +C$ | $-\frac{t_{1,1}}{2} \le t_{\mathrm{M}-1,1/1,0y} - L_1.r_{1,1/1,0z} \le +\frac{t_{1,1}}{2}$<br>$-\frac{t_{1,1}}{2} \le t_{\mathrm{M}-1,1/1,0y} - L_2.r_{1,1/1,0z} \le +\frac{t_{1,1}}{2}$<br>$-C \le t_{\mathrm{M}-1,1/1,0x} \le +C$ |
| 1,2 / 1,0 | $-\frac{t_{1,2}}{2} \le \mathbf{t}_{\mathrm{Q}-1,2/1,0}.\mathbf{x} \le +\frac{t_{1,2}}{2} \perp$<br>$-\frac{t_{1,2}}{2} \le \mathbf{t}_{\mathrm{W}-1,2/1,0}.\mathbf{x} \le +\frac{t_{1,2}}{2}$ | $-C \le \mathbf{t}_{\mathrm{M}-1,2/1,0}.\mathbf{y} \le +C$ | $-\frac{t_{1,2}}{2} \le t_{\mathrm{M}-1,2/1,0x} + L_3.r_{1,2/1,0z} \le +\frac{t_{1,2}}{2}$<br>$-\frac{t_{1,2}}{2} \le t_{\mathrm{M}-1,2/1,0x} + L_5.r_{1,2/1,0z} \le +\frac{t_{1,2}}{2}$<br>$-C \le t_{\mathrm{M}-1,2/1,0y} \le +C$ |
| 1,3 / 1,0 | $-\frac{t_{1,3}}{2} \le \mathbf{t}_{\mathrm{T}-1,3/1,0}.\mathbf{x} \le +\frac{t_{1,3}}{2}$<br>$-\frac{t_{1,3}}{2} \le \mathbf{t}_{\mathrm{V}-1,3/1,0}.\mathbf{x} \le +\frac{t_{1,3}}{2} \perp$ | $-C \le \mathbf{t}_{\mathrm{M}-1,3/1,0}.\mathbf{y} \le +C$ | $-\frac{t_{1,3}}{2} \le t_{\mathrm{M}-1,3/1,0x} + L_3.r_{1,3/1,0z} \le +\frac{t_{1,3}}{2}$<br>$-\frac{t_{1,3}}{2} \le t_{\mathrm{M}-1,3/1,0x} + L_5.r_{1,3/1,0z} \le +\frac{t_{1,3}}{2}$<br>$-C \le t_{\mathrm{M}-1,3/1,0y} \le +C$ |
| 1,4 / 1,0 | $-\frac{t_{1,4}}{2} \le \mathbf{t}_{\mathrm{I}-1,4/1,0}.\mathbf{y} \le +\frac{t_{1,4}}{2}$<br>$-\frac{t_{1,4}}{2} \le \mathbf{t}_{\mathrm{J}-1,4/1,0}.\mathbf{y} \le +\frac{t_{1,4}}{2} \perp$ | $-C \le \mathbf{t}_{\mathrm{M}-1,4/1,0}.\mathbf{x} \le +C$ | $-\frac{t_{1,4}}{2} \le t_{\mathrm{M}-1,4/1,0y} - L_2.r_{1,4/1,0z} \le +\frac{t_{1,4}}{2}$<br>$-\frac{t_{1,4}}{2} \le t_{\mathrm{M}-1,4/1,0y} \le +\frac{t_{1,4}}{2}$<br>$-C \le t_{\mathrm{M}-1,4/1,0x} \le +C$ |
| 2,1 / 2,0 | $-\frac{t_{2,1}}{2} \le \mathbf{t}_{\mathrm{O}-2,1/2,0}.\mathbf{y} \le +\frac{t_{1,1}}{2}$<br>$-\frac{t_{2,1}}{2} \le \mathbf{t}_{\mathrm{P}-2,1/2,0}.\mathbf{y} \le +\frac{t_{1,1}}{2} \perp$ | $-C \le \mathbf{t}_{\mathrm{M}-2,1/2,0}.\mathbf{x} \le +C$ | $-\frac{t_{2,1}}{2} \le t_{\mathrm{M}-2,1/2,0y} - L_1.r_{2,1/2,0z} \le +\frac{t_{2,1}}{2}$<br>$-\frac{t_{2,1}}{2} \le t_{\mathrm{M}-2,1/2,0y} - L_2.r_{2,1/2,0z} \le +\frac{t_{2,1}}{2}$<br>$-C \le t_{\mathrm{M}-2,1/2,0x} \le +C$ |
| 2,2 / 2,0 | $-\frac{t_{2,2}}{2} \le \mathbf{t}_{\mathrm{Y}-2,2/2,0}.\mathbf{x} \le +\frac{t_{2,2}}{2}$<br>$-\frac{t_{2,2}}{2} \le \mathbf{t}_{\mathrm{U}-2,2/2,0}.\mathbf{x} \le +\frac{t_{2,2}}{2} \perp$ | $-C \le \mathbf{t}_{\mathrm{M}-2,2/2,0}.\mathbf{y} \le +C$ | $-\frac{t_{2,2}}{2} \le t_{\mathrm{M}-2,2/2,0x} + L_3.r_{2,2/2,0z} \le +\frac{t_{2,2}}{2}$<br>$-\frac{t_{2,2}}{2} \le t_{\mathrm{M}-2,2/2,0x} + L_5.r_{2,2/2,0z} \le +\frac{t_{2,2}}{2}$<br>$-C \le t_{\mathrm{M}-2,2/2,0y} \le +C$ |
| 2,3 / 2,0 | $-\frac{t_{2,3}}{2} \le \mathbf{t}_{\mathrm{R}-2,3/2,0}.\mathbf{x} \le +\frac{t_{2,3}}{2}$<br>$-\frac{t_{2,3}}{2} \le \mathbf{t}_{\mathrm{S}-2,3/2,0}.\mathbf{x} \le +\frac{t_{2,3}}{2} \perp$ | $-C \le \mathbf{t}_{\mathrm{M}-2,3/2,0}.\mathbf{y} \le +C$ | $-\frac{t_{2,3}}{2} \le t_{\mathrm{M}-2,3/2,0x} + L_3.r_{2,3/2,0z} \le +\frac{t_{2,3}}{2}$<br>$-\frac{t_{2,3}}{2} \le t_{\mathrm{M}-2,3/2,0x} + L_5.r_{2,3/2,0z} \le +\frac{t_{2,3}}{2}$<br>$-C \le t_{\mathrm{M}-2,3/2,0y} \le +C$ |
| 2,4 / 2,0 | $-\frac{t_{2,4}}{2} \le \mathbf{t}_{\mathrm{L}-2,4/2,0}.\mathbf{y} \le +\frac{t_{2,4}}{2} \perp$<br>$-\frac{t_{2,4}}{2} \le \mathbf{t}_{\mathrm{N}-2,4/2,0}.\mathbf{y} \le +\frac{t_{2,4}}{2}$ | $-C \le \mathbf{t}_{\mathrm{M}-2,4/2,0}.\mathbf{x} \le +C$ | $-\frac{t_{2,4}}{2} \le t_{\mathrm{M}-2,4/2,0y} - L_1.r_{2,4/2,0z} \le +\frac{t_{2,4}}{2}$<br>$-\frac{t_{2,4}}{2} \le t_{\mathrm{M}-2,4/2,0y} - L_2.r_{2,4/2,0z} \le +\frac{t_{2,4}}{2}$<br>$-C \le t_{\mathrm{M}-2,4/2,0x} \le +C$ |

Table 3. Definitions of geometric polytopes.

| | Contact constraints | Cap half-spaces | Contact polytopes in M |
|---|---|---|---|
| 2,1 / 1,1 | $\mathbf{t}_{\mathrm{O}-2,1/1,1}.\mathbf{y}=0$<br>$\mathbf{t}_{\mathrm{P}-2,1/1,1}.\mathbf{y}=0$ | $-C\le\mathbf{t}_{\mathrm{M}-2,1/1,1}.\mathbf{x}\le+C$ | $t_{\mathrm{M}-2,1/1,1y}-L_1.r_{2,1/1,1z}=0$<br>$t_{\mathrm{M}-2,1/1,1y}-L_2.r_{2,1/1,1z}=0$<br>$-C\le t_{\mathrm{M}-2,1/1,1x}\le+C$ |
| 2,2 / 1,2 | $0\le-\mathbf{t}_{\mathrm{Q}-2,2/1,2}.\mathbf{x}+d$ | $-\mathbf{t}_{\mathrm{Q}-2,2/1,2}.\mathbf{x}\le+B$<br>$-C\le\mathbf{r}_{2,2/1,2}.\mathbf{z}\le+C$<br>$-C\le\mathbf{t}_{\mathrm{M}-2,2/1,2}.\mathbf{y}\le+C$ | $0\le-\left(t_{\mathrm{M}-2,2/1,2x}+L_3.r_{2,2/1,2z}\right)+d$<br>$-\left(t_{\mathrm{M}-2,2/1,2x}+L_3.r_{2,2/1,2z}\right)+d\le+B$<br>$-C\le\mathbf{r}_{2,2/1,2}.\mathbf{z}\le+C$<br>$-C\le\mathbf{t}_{\mathrm{M}-2,2/1,2}.\mathbf{y}\le+C$ |
| 2,3 / 1,3 | $0\le\mathbf{t}_{\mathrm{T}-2,3/1,3}.\mathbf{x}+d$ | $\mathbf{t}_{\mathrm{T}-2,3/1,3}.\mathbf{x}\le+C$<br>$-C\le\mathbf{r}_{2,3/1,3}.\mathbf{z}\le+C$<br>$-C\le\mathbf{t}_{\mathrm{M}-1,3/1,0}.\mathbf{y}\le+C$ | $0\le t_{\mathrm{M}-2,3/1,3x}+L_3.r_{2,3/1,3z}+d$<br>$t_{\mathrm{M}-2,3/1,3x}+L_3.r_{2,3/1,3z}+d\le+C$<br>$-C\le\mathbf{r}_{2,3/1,3}.\mathbf{z}\le+C$<br>$-C\le\mathbf{t}_{\mathrm{M}-1,3/1,0}.\mathbf{y}\le+C$ |
| 3,4 / 2,4 | $-\frac{J_4}{2}\le\mathbf{t}_{\mathrm{L}-3,4/2,4}.\mathbf{y}\le+\frac{J_4}{2}$<br>$-\frac{J_4}{2}\le\mathbf{t}_{\mathrm{N}-3,4/2,4}.\mathbf{y}\le+\frac{J_4}{2}$ | $-C\le\mathbf{t}_{\mathrm{M}-2,4/3,4}.\mathbf{x}\le+C$ | $-\frac{J_4}{2}\le t_{\mathrm{M}-2,4/3,4y}-L_1.r_{2,4/3,4z}\le+\frac{J_4}{2}$<br>$-\frac{J_4}{2}\le t_{\mathrm{M}-2,4/3,4y}-L_2.r_{2,4/3,4z}\le+\frac{J_4}{2}$<br>$-C\le t_{\mathrm{M}-2,4/3,4x}\le+C$ |

Table 4. Definition of contact polytopes